\documentclass[%
 reprint,
 superscriptaddress,
 nofootinbib,
 amsmath,amssymb,
 aps,
]{revtex4-2}

\usepackage{graphicx}
\usepackage{dcolumn}
\usepackage{bm}
\usepackage{braket}
\usepackage{mathtools}
\usepackage{mathrsfs}

\begin{document}

\title{A quantum algorithm for dynamic mode decomposition\\
       integrated with a quantum differential equation solver}

\author{Yuta Mizuno}
  \email{mizuno@es.hokudai.ac.jp}
  \affiliation{Research Institute for Electronic Science, Hokkaido University,
               Sapporo, Hokkaido 001-0020, Japan}
  \affiliation{Institute for Chemical Reaction Design and Discovery
               (WPI-ICReDD), Hokkaido University, Sapporo, Hokkaido 001-0021, Japan}
  \affiliation{Graduate School of Chemical Sciences and Engineering,
               Hokkaido University, Sapporo, Hokkaido 060-8628, Japan}
\author{Tamiki Komatsuzaki}
  \affiliation{Research Institute for Electronic Science, Hokkaido University,
               Sapporo, Hokkaido 001-0020, Japan}
  \affiliation{Institute for Chemical Reaction Design and Discovery
               (WPI-ICReDD), Hokkaido University, Sapporo, Hokkaido 001-0021, Japan}
  \affiliation{Graduate School of Chemical Sciences and Engineering,
               Hokkaido University, Sapporo, Hokkaido 060-8628, Japan}
  \affiliation{SANKEN, Osaka University, Ibaraki, Osaka 567-0047, Japan}

\date{\today}

\begin{abstract}
  We present a quantum algorithm that analyzes time series data simulated
  by a quantum differential equation solver. The proposed algorithm is
  a quantum version of the dynamic mode decomposition algorithm used in
  diverse fields such as fluid dynamics and epidemiology.
  Our quantum algorithm can also compute matrix eigenvalues and eigenvectors
  by analyzing the corresponding linear dynamical system. Our algorithm handles
  a broad range of matrices, in particular those with complex eigenvalues.
  The complexity of our quantum algorithm is $O(\operatorname{poly}\log N)$
  for an $N$-dimensional system. This is an exponential speedup over known
  classical algorithms with at least $O(N)$ complexity. Thus, our quantum
  algorithm is expected to enable high-dimensional dynamical systems analysis
  and large matrix eigenvalue decomposition, intractable for
  classical computers.
\end{abstract}

\maketitle

\section{Introduction}

Quantum algorithms provide exponential speedups over classical algorithms
for numerical linear algebra tasks such as eigenvalue decomposition of
unitary or Hermitian matrices \cite{Kitaev1995,Cleve1998,Abrams1999},
singular value decomposition of low-rank matrices \cite{Lloyd2014,Rebentrost2018},
and solving linear systems of equations \cite{Harrow2009,Childs2017}.
These quantum algorithms can solve problems of $N$ dimensions in runtime
$O(\operatorname{poly}\log N)$. They have significant applications in quantum
chemistry \cite{Aspuru2005}, machine learning \cite{Lloyd2014,Schuld2016},
and solving differential equations
\cite{Berry2014,Berry2017,Childs2020,Liu2021,Krovi2023,Jennings2024}.

Quantum numerical linear algebra offers prospects for advancements in
dynamical systems analysis. A probability density function on the state space
of a dynamical system is advanced in time by the Perron--Frobenius operator
\cite{Brunton2022,Lin2022}. Meanwhile, the Koopman operator is responsible for
the time evolution of observable functions on the state space
\cite{Brunton2022,Lin2022}. These operators are linear operators on
infinite-dimensional function spaces. In other words,
any finite-dimensional (possibly nonlinear) dynamical system can be
described as an infinite-dimensional linear dynamical system.
Therefore, linear algebraic techniques such as spectral decomposition
can be applied to general dynamical systems analysis.
Such an infinite-dimensional linear operator may be numerically approximated
by a finite-dimensional matrix through a suitable truncated basis expansion.
Although this finite-dimensional approximation may lead to a linear
dynamical system with an extremely-large number of dimensions $N (\gg 1)$,
quantum computing has a potential to handle such high-dimensional systems.

A quantum linear differential equation solver (QLDES) can simulate
an $N$-dimensional linear dynamical system with a sparse coefficient
matrix in runtime $O(\operatorname{poly}\log N)$
\cite{Berry2014,Berry2017,Childs2020,Krovi2023,Jennings2024}.
One can also simulate some nonlinear dynamical systems efficiently
by a QLDES combined with Carleman linearization \cite{Liu2021}, which is
essentially the same as a truncated basis expansion described above.
These quantum solvers yield quantum states whose amplitudes encode
simulated time series data. However, the tomography of such quantum states
takes a runtime of $O(N)$, so that $O(N)$ time and memory space are
required to read out the entire time series data as classical data.
Therefore, an efficient method for extracting essential dynamical
information from the quantum data is highly demanded to achieve
an exponential speedup in dynamical systems analysis \cite{Kiani2022}.

We propose a novel quantum algorithm for dynamic mode decomposition (DMD)
integrated with a QLDES, which achieves an exponential speedup in dynamical
systems analysis over known classical counterparts. The DMD is a numerical
technique that estimates the spectral decomposition of the Koopman operator
of a dynamical system from its time series data \cite{Brunton2022,Tu2014}.
This spectral decomposition elucidates essential temporal behavior of
the dynamical system. Classical DMD algorithms are frequently applied
in various fields such as fluid dynamics and epidemiology \cite{Brunton2022}.
Given that these fields are expected as potential targets of quantum
differential equation analysis \cite{Liu2021,Kiani2022}, our quantum DMD (qDMD)
algorithm should enhance the dynamical systems analysis in these fields
using quantum computers.

Although quantum algorithms for the spectral estimation from time series
data have been proposed by Steffens et al. \cite{Steffens2017} and Xue et al.
\cite{Xue2023}, these algorithms mainly focus on the analysis of time series data
provided as classical data. Because constructing such a classical data
structure requires $O(N)$ time and memory space, they might be unable to
analyze high-dimensional data when $N$ is too large for classical computers
to handle. In contrast, our qDMD algorithm aims to analyze time series data
embedded in quantum states computed by a QLDES. Our framework integrates
simulation and data analysis seamlessly on a quantum computer, achieving
the overall dynamical systems analysis with $O(\operatorname{poly}\log N)$
complexity. A detailed comparison of these algorithms is provided in
Supplemental Material.

Our qDMD algorithm also serves as a quantum subroutine for eigenvalue
decomposition of matrices, especially those with complex eigenvalues
essential in quantum scattering problems \cite{Teplukhin2020}.
If a linear differential equation $\dot{\bm{x}}=\bm{A}\bm{x}$ can be
simulated efficiently on a quantum computer, our algorithm can efficiently
compute approximate eigenvalues and eigenvectors of $\exp(\Delta t \bm{A})$,
where $\Delta t$ is the time step of the simulation. Notably, the matrix
$\bm{A}$ is not restricted to Hermitian and may have complex eigenvalues.
Therefore, the integrated framework of a QLDES and the qDMD algorithm
can be considered as a generalization of quantum phase estimation
\cite{Kitaev1995,Cleve1998,Abrams1999}, which combines Hamiltonian dynamics
simulation and quantum Fourier transform. Although previous studies
\cite{Wang2010,Daskin2014,Teplukhin2020,Shao2022} have pioneered quantum
eigensolvers for complex eigenvalue problems, these approaches have limitations
such as the lack of the theoretical guarantee of an exponential speedup
and requiring a specific form of input states (see Supplemental Material
for details). Our qDMD algorithm is designed to be free from such limitations.

\section{Dynamic Mode Decomposition}

We introduce the \textit{exact DMD} algorithm proposed by Tu et al. \cite{Tu2014}.
Let us consider an $N$-dimensional linear dynamical system
$\dot{\bm{x}}=\bm{A}\bm{x}$, where $\bm{x}\in\mathbb{C}^N$,
and $\bm{A}\in\mathbb{C}^{N \times N}$ is a diagonalizable matrix\footnote{
  For the case that $A$ is not diagonalizable, see discussion
  in Supplemental Material.
}. Let $\bm{K}$ denote the time evolution operator with time step $\Delta t$:
$\bm{K} \coloneqq \exp(\Delta t \bm{A})$. Suppose we have a collection
of $M$ snapshot pairs of time-series data, symbolized as
$\{(\bm{x}_j, \bm{x}_j^\prime)\}_{j=0}^{M-1}$. Here $\bm{x}_j^\prime$
signifies the state observed at the subsequent time step following $\bm{x}_j$:
$\bm{x}_j^\prime \approx \bm{K}\bm{x}_j$ \footnote{
  Since numerical integration of a linear differential equation involves
  approximations, the simulated data $\bm{x}_j^\prime$ is an approximation
  of the exact solution $\bm{K}\bm{x}_j$.
}. Note that $\bm{x}_j$'s can be taken from multiple different trajectories.
From the data, we can estimate the time-evolution operator $\bm{K}$ as
\begin{equation}
\tilde{\bm{K}} =
\operatorname*{argmin}_{\bm{J}\in\mathbb{C}^{N \times N}}
\|\bm{X}^\prime-\bm{J}\bm{X}\|_\mathrm{F}
= \bm{X}^\prime \bm{X}^+,
\end{equation}
where $\tilde{\bm{K}}$ signifies the approximation of the underlying $\bm{K}$,
$\|\cdot\|_\mathrm{F}$ denotes the Frobenius norm,
$\bm{X} \coloneqq [\bm{x}_0 \cdots \bm{x}_{M-1}]$,
$\bm{X}^\prime \coloneqq [\bm{x}_0^\prime \cdots \bm{x}_{M-1}^\prime]$,
and $\bm{X}^+$ is the pseudo-inverse of $\bm{X}$. The construction of
$N \times N$ matrix $\tilde{\bm{K}}$ and its eigenvalue decomposition
becomes intractable as $N$ increases. Thus, we solve
the eigenvalue problem of the following projected matrix instead:
\begin{equation}
\tilde{\bm{K}}^\prime =
\bm{Q}^\dagger \tilde{\bm{K}} \bm{Q},
\end{equation}
where $\bm{Q}$ is an $N \times R$ matrix whose columns are the $R$ dominant
left singular vectors of the $N \times 2M$ matrix $[\bm{X}\ \bm{X}^\prime]$.
The effective rank $R$ is determined so that the error of the rank-$R$
approximation of $[\bm{X}\ \bm{X}^\prime]$ in the Frobenius norm is
less than a specified tolerance. The exact DMD algorithm typically assumes that
$R$ is sufficiently smaller than $N$ so that the eigenvalue decomposition of
the $R \times R$ matrix $\tilde{\bm{K}}^\prime$ is tractable by a (classical)
computer. The eigenvalue decomposition of $\tilde{\bm{K}}^\prime$ approximates
that of $\tilde{\bm{K}}$ as\footnote{
  Without the rank truncation, the eigenvalues and eigenvectors of
  $\tilde{\bm{K}}^\prime$ are exactly those of $\tilde{\bm{K}}$ and
  all of the nonzero eigenvalues of $\tilde{\bm{K}}$ are identified
  by the algorithm \cite{Tu2014}. However, the rank truncation generally
  leads to an approximation error.
}
\begin{equation}
\tilde{\lambda}_r \approx \tilde{\lambda}^\prime_r, \quad
\tilde{\bm{w}}_r \approx \bm{Q}\tilde{\bm{w}}^\prime_r \quad
(r=1, \dots, R).
\end{equation}
Here, $\tilde{\lambda}_r$ and $\tilde{\bm{w}}_r$
(resp. $\tilde{\lambda}_r^\prime$ and $\tilde{\bm{w}}^\prime_r$) are
the $r$-th eigenvalue and eigenvector of $\tilde{\bm{K}}$
(resp. $\tilde{\bm{K}}^\prime$). The real part and the imaginary part
of $(\ln \tilde{\lambda}_r)/\Delta t$ correspond to the decay/growth rate
and the oscillation frequency of the $r$-th DMD mode, respectively.
The computational complexity of this algorithm is $O(\min(NM^2, MN^2))$
for the singular value decomposition (SVD) and $O(R^3)$ for
the eigenvalue decomposition of $\tilde{\bm{K}}^\prime$ \cite{Trefethen1997}.

\section{\lowercase{q}DMD Algorithm}

Our qDMD algorithm consists of the following five steps:
\begin{enumerate}
\item Prepare quantum states encoding $\bm{X}$, $\bm{X}^\prime$, and
      $[\bm{X}\ \bm{X}^\prime]$ using a QLDES.
\item Compute the SVDs of $\bm{X}$, $\bm{X}^\prime$, and
      $[\bm{X}\ \bm{X}^\prime]$ on a quantum computer.
\item Estimate the elements of $\tilde{\bm{K}}^\prime$ from
      the quantum data and construct $\tilde{\bm{K}}^\prime$
      as classical data.
\item Solve the eigenvalue problem of $\tilde{\bm{K}}^\prime$
      on a classical computer.
\item Compute a quantum state encoding $\tilde{\bm{w}}_r$.
\end{enumerate}
Steps 1--3, and 5 are efficiently executed on a quantum computer
in runtime $O(\operatorname{poly}\log N)$ as shown below.
Given that $R \ll N$, step 4 can be handled by a classical computer.
Consequently, our qDMD algorithm is exponentially faster than
its classical counterpart with respect to $N$. Similar quantum-classical
hybrid strategies are also employed by Steffens et al. \cite{Steffens2017} and
Xue et al. \cite{Xue2023}, though their methods perform steps 1 and 2 using
quantum oracles of classical time series data and the specifics of
the quantum procedures differ from our algorithm
(see also Supplemental Material for algorithm comparison).

In what follows, we will expound the quantum procedures of steps 1--3 and 5.
Henceforth, we adopt the following notation: The computational basis
whose bit string represents integer $i$ is denoted by $\ket{i}$. As necessary,
we denote a ket vector of the $k$-th quantum register like $\ket{\ }_k$.
For vector $\bm{v} = (v^0, \cdots, v^{n-1})^\top \in \mathbb{C}^n$, we define
$\ket{\bm{v}} \coloneqq \sum_{i=0}^{n-1} v^i\ket{i}$ (unnormalized).
Similarly, for matrix
$\bm{Z} = [\bm{v}_0 \cdots \bm{v}_{m-1}] \in \mathbb{C}^{n \times m}$,
we write
$\ket{\bm{Z}} \coloneqq \sum_{j=0}^{m-1} \ket{\bm{v}_j}\!\ket{j}
 = \sum_{i=0}^{n-1}\sum_{j=0}^{m-1}v_j^i\ket{i}\!\ket{j}$.
A normalized matrix $\bm{Z}/\|\bm{Z}\|_\mathrm{F}$ is denoted by
$\hat{\bm{Z}}$, thus $\ket{\hat{\bm{Z}}}$ symbolizes the normalized ket vector
(quantum state) proportional to $\ket{\bm{Z}}$. Additionally, the $r$-th
singular value, left and right singular vectors of matrix $\bm{Z}$ are
designated by $\sigma^{\tiny \bm{Z}}_r$, $\bm{u}^{\tiny \bm{Z}}_r$, and
$\bm{v}^{\tiny \bm{Z}}_r$, respectively. The notation of quantum circuit
diagrams we employ can be found in \cite{Nielsen2010}.

\subsection{Step 1}

First, we prepare a quantum state encoding matrix $\bm{X}$. In this article,
we consider preparing time series data of $L$ different trajectories of
$T$ time steps, thus the number of columns $M$ equals $TL$.
In addition, we normalize the time scale in units of the simulation time step
$\Delta t \le 1/\|\bm{A}\|_2$ ($\|\cdot\|_2$ denotes the spectral norm)
for convenience. We assume a quantum oracle $U_0$ that generates
a superposition of $L$ initial states $\{\bm{x}_k\}_{k=0}^{L-1}$ as
\begin{equation}
U_0 \ket{0} \!\ket{0}
= \frac{1}{\sqrt{\sum_k \|\bm{x}_k\|^2}}
  \sum_{k=0}^{L-1} \ket{\bm{x}_k} \!\ket{k}.
\end{equation}
We also introduce a block encoding of matrix $\bm{A}$, which is
a unitary operator $U_{\bm{A}}$ that satisfies
\begin{equation}
U_{\bm{A}} \ket{\bm{x}} \! \ket{0}_\mathrm{a}
= \omega^{-1}\ket{\bm{A}\bm{x}} \! \ket{0}_\mathrm{a} + \ket{\perp}
\end{equation}
for an arbitrary $N$-dimensional vector $\bm{x}$. Here, $\ket{\ }_\mathrm{a}$
denotes a ket of an ancilla register,  $\omega$ is a rescaling factor,
and $(I\otimes\ket{0}_\mathrm{a}\!\bra{0}_\mathrm{a})\ket{\perp} = 0$.
We further assume that the computational costs of these operators are
$O(\operatorname{poly}\log NL)$. Using $U_0$ and $U_{\bm{A}}$,
we prepare a quantum state proportional to $\ket{\bm{X}}$ via a QLDES
\cite{Berry2017,Krovi2023,Jennings2024}. The ket $\ket{\bm{X}}$ is given by
\begin{equation}
\ket{\bm{X}}
= \sum_{k=0}^{L-1} \sum_{t=0}^{T-1} \ket{\bm{x}_k(t)}_1 \!\ket{k}_2 \!\ket{t}_3,
\end{equation}
where $\bm{x}_k(t)$ denotes the state at the $t$-th time step of the trajectory
initiated from $\bm{x}_k$. The first register encodes states of the dynamical
system, and the second and third registers---indicating the initial condition
$k$ and the time step count $t$---collectively label the column index of
$\bm{X}$ as $\ket{kT+t}_{23} = \ket{k}_2\!\ket{t}_3$. The gate complexity
of the QLDES is $O(T\operatorname{poly}\log(NM/\epsilon))$ for tolerant
simulation error $\epsilon$.

Next, a quantum state encoding $\bm{X}^\prime$ is prepared by one-time-step
simulation of $\dot{\bm{x}}=\bm{A}\bm{x}$ by the QLDES. By treating the $M$
columns of $\bm{X}$ as initial states and adding a qubit as the fourth register
to indicate the time step count, the QLDES computes a quantum state
proportional to
\begin{equation}
\ket{[\bm{X}\ \bm{X}^\prime]} =
\ket{\bm{X}} \!\ket{0}_4 + \ket{\bm{X}^\prime} \!\ket{1}_4
\end{equation}
with gate complexity $O(\operatorname{poly}\log(NM/\epsilon))$ \footnote{
  In the one-step simulation, access to a unitary oracle that generates
  the initial states $\ket{\hat{\bm{X}}}$ is not available; however we can
  still prepare multiple copies of $\ket{\hat{\bm{X}}}$. In this situation,
  we may not be able to apply the amplitude amplification protocol
  \cite{Childs2017} to reduce the complexity with respect to the condition number
  of an underlying linear system in the QLDES routine. The condition number is
  $\tilde{O}(T \sup_{t \in [0,T-1]}\|\exp(t\bm{A})\|_2)$
  \cite{Krovi2023,Jennings2024}, which is $O(1)$ for the one-step simulation.
  Therefore, the inaccessibility of the quantum oracle generating
  $\ket{\hat{\bm{X}}}$ should not be a critical issue.
}. This ket vector can be viewed as encoding $[\bm{X}\ \bm{X}^\prime]$,
regarding the second to fourth resisters as indicating the column index
collectively. Measuring the fourth register, we obtain a quantum state
$\ket{\hat{\bm{X}}}$ or $\ket{\hat{\bm{X}}^\prime}$.

\subsection{Step 2}

According to the procedure proposed by Schuld et al. \cite{Schuld2016},
we perform the SVD of a normalized matrix $\hat{\bm{Z}}$
($\bm{Z}=\bm{X}, \bm{X}^\prime,\ \text{or}\ [\bm{X}\ \bm{X}^\prime]$)
on a quantum computer using $C$ copies of $\ket{\hat{\bm{Z}}}$ as
\begin{equation}
\ket{\hat{\bm{Z}}}^{\!\otimes C}
\mapsto
\ket{\mathrm{SVD}(\hat{\bm{Z}})} \approx 
\sum_{r=1}^R \hat{\sigma}^{\tiny \bm{Z}}_r
\!\ket{\bm{u}^{\tiny \bm{Z}}_r} \!\ket{{\bm{v}^{\tiny \bm{Z}}_r}^*}
\!\ket{(\hat{\sigma}^{\tiny \bm{Z}}_r)^2}_5,
\end{equation}
where
$\hat{\sigma}^{\tiny \bm{Z}}_r \coloneqq \sigma^{\tiny \hat{\bm{Z}}}_r =
 \sigma^{\tiny \bm{Z}}_r/\|\bm{Z}\|_\mathrm{F}$, and
$\ket{(\hat{\sigma}^{\tiny \bm{Z}}_r)^2}_5$ designates
the computational basis of the extra fifth register indicating
the binary representation of $(\hat{\sigma}^{\tiny \bm{Z}}_r)^2$.
Note that matrix normalization does not change singular vectors:
$\bm{u}^{\tiny \hat{\bm{Z}}}_r = \bm{u}^{\tiny \bm{Z}}_r$ and
$\bm{v}^{\tiny \hat{\bm{Z}}}_r = \bm{v}^{\tiny \bm{Z}}_r$.
Thus we omit the hat ($\hat{\ }$) in the superscript of singular vectors
for brevity. This quantum SVD process utilizes density matrix exponentiation
\cite{Lloyd2014} and quantum phase estimation. The necessary number of state
copies $C$ for precision $\epsilon$ is $O(1/\epsilon^2)$ \cite{Kimmel2017}.

\subsection{Step 3}

The estimation of $\tilde{\bm{K}}^\prime$ is based on
the following factorization:
\begin{equation}
\tilde{\bm{K}}^\prime \approx
\frac{\|\bm{X}^\prime\|_\mathrm{F}}{\|\bm{X}\|_\mathrm{F}}
(\bm{Q}^\dagger \bm{U}^\prime) \hat{\bm{\Sigma}}^\prime
(\bm{V}^{\prime \dagger} \bm{V}) \hat{\bm{\Sigma}}^{-1}
(\bm{U}^\dagger \bm{Q}),
\end{equation}
where
$\hat{\bm{X}} \approx
 \bm{U} \hat{\bm{\Sigma}} \bm{V}^\dagger$ and
$\hat{\bm{X}}^\prime \approx
 \bm{U}^\prime \hat{\bm{\Sigma}}^\prime \bm{V}^{\prime \dagger}$
are the SVDs of the normalized data matrices with rank-$R$ truncation.
The first factor
$\|\bm{X}^\prime\|_\mathrm{F}/\|\bm{X}\|_\mathrm{F}
 ~(=\|\!\ket{\bm{X}^\prime}\!\|/\|\!\ket{\bm{X}}\!\|)$
can be estimated by measuring the fourth register of
$\ket{[\bm{X}\ \bm{X}^\prime]}$ because the probability ratio of measured
values $1$ to $0$ equals the square of this factor. The diagonal elements
of $\hat{\bm{\Sigma}}$ and $\hat{\bm{\Sigma}}^\prime$,
i.e., $\{\hat{\sigma}^{\tiny \bm{X}}_r\}_{r=1}^R$ and
$\{\hat{\sigma}^{\tiny \bm{X}^\prime}_r\}_{r=1}^R$,
can be estimated by measuring the fifth register of
$\ket{\operatorname{SVD}(\hat{\bm{X}})}$ and
$\ket{\operatorname{SVD}(\hat{\bm{X}}^\prime)}$. All the off-diagonal
elements of $\hat{\bm{\Sigma}}$ and $\hat{\bm{\Sigma}}^\prime$ are zero.
The elements of matrices $\bm{Q}^\dagger \bm{U}^\prime$,
$\bm{U}^\dagger \bm{Q}$, and $\bm{V}^{\prime \dagger} \bm{V}$
are inner products between singular vectors. Note that the $r$-th
column vector of $\bm{Q}$ corresponds to
$\bm{u}^{\tiny [\bm{X}\ \bm{X}^\prime]}_r$.
Now, the remaining task is to estimate
$\braket{\bm{u}^{\tiny [\bm{X}\ \bm{X}^\prime]}_r|
         \bm{u}^{\tiny \bm{X}^\prime}_{r^\prime}}$,
$\braket{\bm{u}^{\tiny \bm{X}}_r|
         \bm{u}^{\tiny [\bm{X}\ \bm{X}^\prime]}_{r^\prime}}$, and
$\braket{\bm{v}^{\tiny \bm{X}^\prime}_r|\bm{v}^{\tiny \bm{X}}_{r^\prime}}$
for $R^2$ combinations of $r$ and $r^\prime$.

\begin{figure}[!tb]
    \centering
    \includegraphics{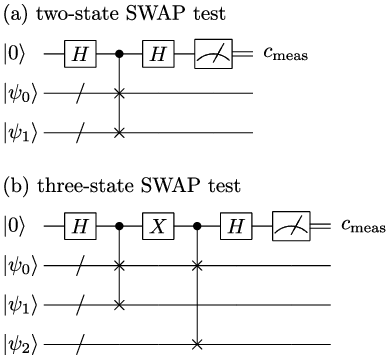}
    \caption{
      Quantum circuit for inner product estimation using controlled SWAP gates.
      The input states $\ket{\psi_k}$ ($k=0, 1, 2$) are arbitrary quantum states.
      (a) In the two-state SWAP test,
          $\operatorname{Pr}[c_\mathrm{meas}=0] -
           \operatorname{Pr}[c_\mathrm{meas}=1] =
           |\!\braket{\psi_0|\psi_1}\!|^2$.
      (b) In the three-state SWAP test,
          $\operatorname{Pr}[c_\mathrm{meas}=0] -
           \operatorname{Pr}[c_\mathrm{meas}=1] =
           \operatorname{Re}(\braket{\psi_0|\psi_1}
                            \!\braket{\psi_1|\psi_2}
                            \!\braket{\psi_2|\psi_0})$.
          When a phase gate $S$ is applied to the first qubit just after
          the first Hadamard gate,
          $\operatorname{Pr}[c_\mathrm{meas}=0] -
           \operatorname{Pr}[c_\mathrm{meas}=1] =
           \operatorname{Im}(\braket{\psi_0|\psi_1}
                            \!\braket{\psi_1|\psi_2}
                            \!\braket{\psi_2|\psi_0})$.
    }
    \label{fig:swap-test}
\end{figure}

The two-state SWAP test depicted in Fig.~\ref{fig:swap-test}~\!(a) is often employed for
estimating the absolute value of the inner product between arbitrary quantum
states $\ket{\psi_0}$ and $\ket{\psi_1}$. However, the two-state SWAP test
cannot estimate the phase (argument) of the inner product. Furthermore,
the global phase of a singular vector is arbitrary. For instance,
if we have a singular vector pair
$(\ket{\bm{u}_r}, \ket{\bm{v}^*_r})$, then
$(\mathrm{e}^{\mathrm{i}\theta}\!\ket{\bm{u}_r},
  \mathrm{e}^{-\mathrm{i}\theta}\!\ket{\bm{v}^*_r})$
is also a valid pair, where $\theta$ ranges from 0 to $2\pi$. The choice of
the global phase of the singular vector pair changes inner products to be
estimated. To overcome these challenges, we introduce
the \textit{three-state SWAP test} (Fig.~\ref{fig:swap-test}~\!(b)) and
\textit{reference states} for the left and right singular vectors.

First, we estimate the inner products between left singular vectors.
We define the global phase of each left singular vector state
$\ket{\bm{u}}$ such that $\arg\braket{\chi_1|\bm{u}}=0$ for a fixed
reference quantum state $\ket{\chi_1}$. The choice of $\ket{\chi_1}$
is arbitrary\footnote{
  However, the choice of $\ket{\chi_1}$ and $\ket{\chi_2}$ affects
  the algorithm's efficiency (see Supplemental Material).
  \label{foot:reference-state}
}, provided that $\braket{\chi_1|\bm{u}} \neq 0$
for all relevant left singular vectors $\bm{u}$ and that it can be
prepared with $O(\operatorname{poly}\log N)$ complexity by a unitary
gate as $U_{\chi_1}\ket{0} = \ket{\chi_1}$. The two-state SWAP test
between $\ket{\chi_1}$ and $\ket{\bm{u}}$ estimates
$|\!\braket{\chi_1|\bm{u}}\!|$. Here, the singular vector state
$\ket{\bm{u}}$ can be prepared by executing the quantum SVD and
measuring the fifth register encoding squared singular values.
Additionally, the three-state SWAP test between $\ket{\chi_1}$ and arbitrary
left singular vector states $\ket{\bm{u}}$ and $\ket{\bm{u}^\prime}$
provides an estimate of
$\braket{\chi_1|\bm{u}}\!\braket{\bm{u}|\bm{u}^\prime}\!
 \braket{\bm{u}^\prime|\chi_1}$.
Leveraging the known absolute values and phases of $\braket{\chi_1|\bm{u}}$
and $\braket{\bm{u}^\prime|\chi_1}$, we can derive an estimate
of $\braket{\bm{u}|\bm{u}^\prime}$. In this way,
$\braket{\bm{u}^{\tiny [\bm{X}\ \bm{X}^\prime]}_r|
         \bm{u}^{\tiny \bm{X}^\prime}_{r^\prime}}$ and
$\braket{\bm{u}^{\tiny \bm{X}}_r|
         \bm{u}^{\tiny [\bm{X}\ \bm{X}^\prime]}_{r^\prime}}$
can be estimated.

\begin{figure}[!tb]
    \centering
    \includegraphics{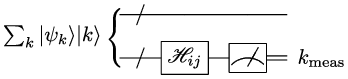}
    \caption{
      Quantum circuit for inner product estimation using a Hadamard gate.
      The $\ket{\psi_k}$'s are arbitrary ket vectors that satisfy
      $\sum_k \|\!\ket{\psi_k}\!\|^2 = 1$. Gate $\mathscr{H}_{ij}$ is defined by
      $H_{ij} + I^\perp_{ij}$. Here, $H_{ij}$ is the Hadamard gate on
      the 2-dimensional Hilbert space
      $\mathcal{H}\coloneqq\operatorname{span}\{\ket{i}, \ket{j}\}$, i.e.,
      $H_{ij}\coloneqq(\ket{i}\!\bra{i}+\ket{i}\!\bra{j}
                      +\ket{j}\!\bra{i}-\ket{j}\!\bra{j})/\!\sqrt{2}$,
      and $I^\perp_{ij}$ is the identity operator on the orthogonal
      complementary space of $\mathcal{H}$. The measurement probability satisfies
      $\operatorname{Pr}[k_\mathrm{meas}=i]-
       \operatorname{Pr}[k_\mathrm{meas}=j] =
       2\operatorname{Re}\braket{\psi_i | \psi_j}$.
      When the phase shift $\ket{j} \mapsto -\mathrm{i}\ket{j}$ is applied
      before $\mathscr{H}_{ij}$,
      $\operatorname{Pr}[k_\mathrm{meas}=i]-
       \operatorname{Pr}[k_\mathrm{meas}=j] =
       2\operatorname{Im}\braket{\psi_i | \psi_j}$.
    }
    \label{fig:hadamard-inner-product-estimation}
\end{figure}

Next, we estimate the inner products between right singular vectors.
As with $\ket{\chi_1}$, a reference state $\ket{\chi_2}$ for
right singular vectors can be chosen arbitrarily\footref{foot:reference-state},
provided that $\braket{\chi_2|\bm{v}^*} \neq 0$ for all relevant right singular
vectors $\bm{v}$ and that it can be prepared with $O(\operatorname{poly}\log N)$
complexity by a unitary gate as $U_{\chi_2}\ket{0} = \ket{\chi_2}$.
Since the global phase of a right singular vector is synchronized with
its corresponding left singular vector, we cannot arbitrarily define
$\arg\braket{\chi_2|\bm{v}^*}$; instead, we need to estimate it as well.
Once we determine $\braket{\chi_2|\bm{v}^*}$ for every right singular
vector $\bm{v}$, we can estimate
$\braket{\bm{v}^{\tiny \bm{X}^\prime}_r|\bm{v}^{\tiny \bm{X}}_{r^\prime}}$
using the three-state SWAP test as described above.
Thus, let us consider how to determine $\braket{\chi_2|\bm{v}^*}$.
We employ the quantum circuit shown in
Fig.~\ref{fig:hadamard-inner-product-estimation}
for this inner product estimation. The input state to this circuit is
prepared according to the following protocol:
\begin{enumerate}
\item  Initialize the first to fifth registers as in steps 1 and 2
       of the qDMD algorithm. Additionally, append a sixth register initialized
       with a single qubit in the state $(\ket{0}_6+\ket{1}_6)/\sqrt{2}$.
\item  Perform linear differential equation simulation
       to create
       $\ket{\hat{[\bm{X}\ \bm{X}^\prime]}}$,
       conditioned on the sixth register being zero.
\item  Perform the quantum SVD of $\hat{\bm{X}}$,
       conditioned on both the fourth and sixth registers being zero.
\item  Apply the unitary gates that create the reference states as
       $(U_{\chi_1} \otimes U_{\chi_2})\ket{0}_1\!\ket{0}_{23} =
       \ket{\chi_1}_1\!\ket{\chi_2}_{23}$, conditioned on
       the sixth register being one.
\end{enumerate}
This protocol yields the quantum state given by
\begin{widetext}
\begin{equation}
\frac{1}{\sqrt{2}\|[\bm{X}\ \bm{X}^\prime]\|_\mathrm{F}}
\left[
  \|\bm{X}\|_\mathrm{F} \sum_{r=1}^R
  \hat{\sigma}^{\tiny \bm{X}}_r \ket{\bm{u}^{\tiny \bm{X}}_r}_1\!
  \ket{\bm{v}^{\tiny \bm{X}*}_r}_{23}\! \ket{0}_4\!
  \ket{(\hat{\sigma}^{\tiny \bm{X}}_r)^2}_5 +
  \ket{\bm{X}^\prime}_{123}\! \ket{1}_4\! \ket{0}_5
\right]\ket{0}_6
+ \frac{1}{\sqrt{2}}
\ket{\chi_1}_1\! \ket{\chi_2}_{23}\!
\ket{0}_4\! \ket{0}_5\! \ket{1}_6.
\end{equation}
\end{widetext}
Let us set $\ket{0}_4\!\ket{0}_5\!\ket{1}_6$
and $\ket{0}_4\!\ket{(\hat{\sigma}^{\tiny \bm{X}}_r)^2}_5\!\ket{0}_6$ to
$\ket{i}$ and $\ket{j}$ in the circuit diagram shown in
Fig.~\ref{fig:hadamard-inner-product-estimation}, respectively.
Then, the circuit provides an estimate of
$\braket{\chi_1|\bm{u}^{\tiny \bm{X}}_r}\!
 \braket{\chi_2|\bm{v}^{\tiny \bm{X} *}_r}$.
Since we know the value of $\braket{\chi_1|\bm{u}^{\tiny \bm{X}}_r}$,
we can derive an estimate of $\braket{\chi_2|\bm{v}^{\tiny \bm{X} *}_r}$.
Similarly, we can estimate $\braket{\chi_2|\bm{v}^{\tiny \bm{X}^\prime *}_r}$
by performing the quantum SVD of $\bm{X}^\prime$, conditioned on
the fourth register being one instead of zero, in the third step of
the above protocol.

The number of quantum SVDs necessary for estimating $\tilde{\bm{K}}^\prime$
with precision $\epsilon$ is $O(1/\epsilon^2 \operatorname{poly}R)$,
excluding reference state preparation costs. The factor $O(1/\epsilon^2)$
originates from sampling errors obeying the central limit theorem.
While preparing the reference states may require additional
quantum SVDs, the overall gate complexity remains at
$O(\operatorname{poly}\log N)$. A detailed analysis of
the computational complexity can be found in Supplemental Material.

\subsection{Step 5}

\begin{figure}[!tb]
    \centering
    \includegraphics{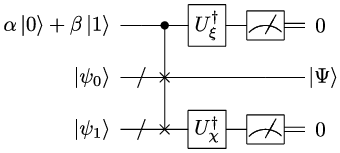}
    \caption{
      Quantum circuit for coherent addition of two quantum states
      $\ket{\psi_0}$ and $\ket{\psi_1}$ \cite{Oszmaniec2016}.
      $U_\chi$ is a unitary gate satisfying $U_\chi\ket{0}=\ket{\chi}$.
      $U_\xi$ is a unitary gate satisfying $U_\xi\ket{0}=\ket{\xi}$,
      where
      $\ket{\xi} \coloneqq
       \sqrt{c_0/(c_0+c_1)}\ket{0}+\sqrt{c_1/(c_0+c_1)}\ket{1}$.
      If the measured values of the first and third registers are both zero,
      the coherent addition has been successful.
    }
    \label{fig:coherent-state-addition}
\end{figure}

A quantum state encoding the $r$-th DMD mode is given by
\begin{equation}
\ket{\tilde{\bm{w}}_r} \approx
\sum_{r^\prime=1}^R \tilde{w}_r^{\prime r^\prime}
\ket{\bm{u}^{\tiny [\bm{X}\ \bm{X}^\prime]}_{r^\prime}},
\end{equation}
where
$\tilde{\bm{w}}_r^\prime =
 (\tilde{w}_r^{\prime 1}, \dots, \tilde{w}_r^{\prime R})^\top$
is computed at step 4.
Such a coherent superposition of quantum states can be created using
the quantum circuit shown in Fig.~\ref{fig:coherent-state-addition}.
This circuit creates a superposition of $\ket{\psi_0}$ and $\ket{\psi_1}$
\cite{Oszmaniec2016}:
\begin{equation}
\ket{\Psi} =
\alpha\frac{\braket{\chi|\psi_1}}{|\!\braket{\chi|\psi_1}\!|}\ket{\psi_0} +
\beta \frac{\braket{\chi|\psi_0}}{|\!\braket{\chi|\psi_0}\!|}\ket{\psi_1}.
\end{equation}
Here, $\alpha$ and $\beta$ are user-specified complex amplitudes,
and $\ket{\chi}$ is a reference quantum state. This addition process is
probabilistic. The success probability is $c_0c_1/(c_0+c_1)$
if $\braket{\psi_0|\psi_1}=0$, where $c_i = |\!\braket{\chi|\psi_i}\!|^2$.
By recursively creating coherent superpositions of two states,
we can construct the multi-state superposition $\ket{\tilde{\bm{w}}_r}$
with $O(\operatorname{poly} R)$ times of the quantum SVD
(see Supplemental Material).

\section{Conclusion}

The qDMD algorithm performs DMD on quantum time series data generated
by a QLDES. This algorithm is also capable of computing (possibly complex)
eigenvalues and eigenvectors of matrices. Excluding reference state
preparation costs, the total gate complexity scales as
$O(T\operatorname{poly}\log(NM/\epsilon)\operatorname{poly}(R)/\epsilon^4)$.
The qDMD algorithm can achieve an exponential speedup over its classical
counterpart in terms of $N$ if $R$ remains at most
$O(\operatorname{poly}\log N)$. Since the algorithm utilizes density matrix
exponentiation and sampling-based inner product estimation, the dependency on
$\epsilon$ is less optimal than that of the classical counterpart.
Reducing the complexity with respect to $\epsilon$ should be addressed
in future work.

\begin{acknowledgments}
This work was supported by JST, PRESTO Grant Number JPMJPR2018, Japan,
and partially by ``Crossover Alliance to Create the Future with People,
Intelligence and Materials" from MEXT, Japan (to YM).
\end{acknowledgments}

\bibliography{main}

\end{document}


\title{Supplemental Material for\\
       ``A quantum algorithm for dynamic mode decomposition\\
       integrated with a quantum differential equation solver"}

\author{Yuta Mizuno}
  \email{mizuno@es.hokudai.ac.jp}
  \affiliation{Research Institute for Electronic Science, Hokkaido University,
               Sapporo, Hokkaido 001-0020, Japan}
  \affiliation{Institute for Chemical Reaction Design and Discovery
               (WPI-ICReDD), Hokkaido University, Sapporo, Hokkaido 001-0021, Japan}
  \affiliation{Graduate School of Chemical Sciences and Engineering,
               Hokkaido University, Sapporo, Hokkaido 060-8628, Japan}
\author{Tamiki Komatsuzaki}
  \affiliation{Research Institute for Electronic Science, Hokkaido University,
               Sapporo, Hokkaido 001-0020, Japan}
  \affiliation{Institute for Chemical Reaction Design and Discovery
               (WPI-ICReDD), Hokkaido University, Sapporo, Hokkaido 001-0021, Japan}
  \affiliation{Graduate School of Chemical Sciences and Engineering,
               Hokkaido University, Sapporo, Hokkaido 060-8628, Japan}
  \affiliation{SANKEN, Osaka University, Ibaraki, Osaka 567-0047, Japan}

\date{\today}

\maketitle

\tableofcontents

\section{Computational Complexity of Estimating $\tilde{\bm{K}}^\prime$}

This section offers a detailed analysis of the computational complexity
involved in estimating $\tilde{\bm{K}}^\prime$. We assume that the required
precision of each element of $\tilde{\bm{K}}^\prime$ is $\epsilon$ in this
section. Note that we omit the discussion on the computational complexity
of preparing reference states for the inner product estimation, which will be
presented in the next section.

In what follows, we first evaluate the required number of repetitions of
each elemental estimation process, such as the three-state SWAP test for
estimating an inner product. Next, we evaluate the required number of
quantum SVDs for collecting singular values and their associated singular
vector states necessary for the singular value estimation and the two-state
and three-state SWAP tests. A single quantum SVD process yields
a quantum state $\ket{\mathrm{SVD}(\hat{\bm{Z}})}$ and the measurement
of the fifth register encoding singular values provides a triple of
a singular value and its associated left and right singular vector states
\textit{at random}. Therefore, we need to take into account this probabilistic
nature to estimate the required number of quantum SVDs. To this end,
we introduce a general theorem on the necessary number of trials of
random selection, proven by Brayton \cite{Brayton1963}. Finally, we evaluate
the total number of quantum SVDs necessary for estimating the whole matrix
$\tilde{\bm{K}}^\prime$ as well as its gate complexity.

\subsection{Required number of repetitions of each elemental estimation process}

\begin{table*}[tb]
  \centering
  \caption{
   Summary of the elemental estimation processes
   in estimating $\tilde{\bm{K}}^\prime$.
  }
  {\renewcommand\arraystretch{2}
  \begin{tabular}{l|l|p{6cm}|l}
    \hline
    Estimation Target & Input State(s) & Estimation Method & Required Number of Repetitions \\
    \hline
    $\frac{\|\bm{X}^\prime\|_\mathrm{F}}{\|\bm{X}\|_\mathrm{F}}$ &
    $\ket{[\bm{X}\ \bm{X}^\prime]}_{1:4}$ &
    Measuring the 4th register & $O\left(\frac{1}{\epsilon^2}\right)$ \\
    $\hat{\sigma}$ & $\ket{\hat{\sigma}^2}$ &
    Readout of the bit string of the register & $1$ \\
    $\braket{\chi_1|\bm{u}}$ & $\ket{\chi_1}, \ket{\bm{u}}$ &
    Two-state SWAP test & $O\left(\frac{1}{\epsilon^2}\right)$ \\
    $\braket{\bm{u}|\bm{u}^\prime}$ & $\ket{\chi_1}, \ket{\bm{u}}, \ket{\bm{u}^\prime}$ &
    Three-state SWAP test &
    $O\left(\frac{1}{|\braket{\chi_1|\bm{u}}|^2|
                      \braket{\chi_1|\bm{u}^\prime}|^2\epsilon^2}\right)$ \\
    $\braket{\bm{v}|\bm{v}^\prime}$ &$\ket{\chi_2}, \ket{\bm{v}^*}, \ket{\bm{v}^{\prime *}}$ &
    Three-state SWAP test &
    $O\left(\frac{1}{|\braket{\chi_2|\bm{v}^*}|^2
                     |\braket{\chi_2|\bm{v}^{\prime *}}|^2\epsilon^2}\right)$ \\
    $\braket{\chi_2|\bm{v}^{\tiny \bm{Z}*}_r}$ &
    $\ket{0}_{1:5}\frac{\ket{0}_6+\ket{1}_6}{\sqrt{2}}$ &
    The process utilizing the circuit shown in Fig. 2 of the main text &
    $O\left(\frac{1}{P_{\tiny \bm{Z}}(\hat{\sigma}^{\tiny \bm{Z}}_r)^2
                     |\braket{\chi_1|\bm{u}^{\tiny \bm{Z}}_r}|^2\epsilon^2}\right)$ \\
    \hline
    \end{tabular}}
  \label{tbl:estimation-processes}
\end{table*}

Table~\ref{tbl:estimation-processes} summarizes the elemental estimation process
and their required numbers of repetitions.

The factor $\|\bm{X}^\prime\|_\mathrm{F}/\|\bm{X}\|_\mathrm{F}$ is
estimated through the measurement of the fourth register of
$\ket{[\bm{X}\ \bm{X}^\prime]}_{1:4}$. Here, we denote a ket vector of
the $k$-th to $k^\prime$-th registers like $\ket{\ }_{k:k^\prime}$.
This estimation relies on the estimation of the probability of measuring
zero and the probability of measuring one. The error of the probability
estimation decrease as the number of samples increases obeying
the central limit theorem. Therefore, $O(1/\epsilon^2)$ repetitions
of the estimation process is needed to estimate the factor
with precision $\epsilon$.

Each normalized singular value $\hat{\sigma}$ of matrix $\bm{Z}$ is
encoded in the fifth register of $\ket{\mathrm{SVD}(\hat{\bm{Z}})}_{1:5}$.
Measuring the fifth register, the quantum state of the fifth register
collapses into a computational basis state that encodes one of
the normalized singular values of $\bm{Z}$ as a bit string. Consequently,
a single readout of the bit string suffices to estimate the singular value.
Note that it is probabilistic which singular value is measured;
this probabilistic nature will be considered in the next subsection.

For each left singular vector $\bm{u}$, the inner product
$\braket{\chi_1|\bm{u}}$ is estimated by the two-state SWAP test.
The two-state SWAP test estimates the inner product through measurement
probability estimation. Thus $O(1/\epsilon^2)$ repetitions of the two-state
SWAP test is necessary for estimating $\braket{\chi_1|\bm{u}}$ with precision
$\epsilon$.

For each pair of left singular vectors $\bm{u}$ and $\bm{u}^\prime$,
the inner product $\braket{\bm{u}|\bm{u}^\prime}$ is estimated by
the three-state SWAP test. The three-state SWAP test estimates
$\braket{\chi_1|\bm{u}}\!\braket{\bm{u}|\bm{u}^\prime}\!
 \braket{\bm{u}^\prime|\chi_1}$.
Therefore, the required precision of the three-state SWAP test is
$|\!\braket{\chi_1|\bm{u}}\!||\!\braket{\chi_1|\bm{u}^\prime}\!|\epsilon$
for estimating $\braket{\bm{u}|\bm{u}^\prime}$ with precision $\epsilon$.
Given that the estimation error obeys the central limit theorem,
the required number of the three-state SWAP test for estimating
$\braket{\bm{u}|\bm{u}^\prime}$ is
$O(1/|\!\braket{\chi_1|\bm{u}}\!|^2
     |\!\braket{\chi_1|\bm{u}^\prime}\!|^2\epsilon^2)$.
Likewise, for each pair of right singular vectors $\bm{v}$ and $\bm{v}^\prime$,
the estimation of $\braket{\bm{v}|\bm{v}^\prime}$ requires
$O(1/|\!\braket{\chi_2|\bm{v}^*}\!|^2
     |\!\braket{\chi_2|\bm{v}^{\prime *}}\!|^2\epsilon^2)$
repetitions of the three-state SWAP test.

For each right singular vector $\bm{v}^{\tiny \bm{Z}}_r$
($\bm{Z}=\bm{X}, \bm{X}^\prime$, $r=1, \dots, R$),
the inner product $\braket{\chi_2|\bm{v}^{\tiny \bm{Z}*}_r}$ is
estimated by the process utilizing the circuit depicted in Fig.~2
of the main text. This process estimate
$\sqrt{P_{\tiny \bm{Z}}}\hat{\sigma}^{\tiny \bm{Z}}_r\!
 \braket{\chi_1|\bm{u}^{\tiny \bm{Z}}_r}\!
 \braket{\chi_2|\bm{v}^{\tiny \bm{Z} *}_r}$ through probability estimation.
Here, $P_{\tiny \bm{Z}}$ denotes the probability of obtaining $\ket{\bm{Z}}$
by measuring the fourth register of $\ket{[\bm{X}\ \bm{X}^\prime]}$, that is,
\begin{equation}
P_{\tiny \bm{Z}} \coloneqq
\frac{\|\bm{Z}\|_\mathrm{F}^2}{\|[\bm{X}\ \bm{X}^\prime]\|_\mathrm{F}^2}.
\end{equation}
Consequently, the required number of repetitions of running the process is
$O(1/P_{\tiny \bm{Z}}(\hat{\sigma}^{\tiny \bm{Z}}_r)^2
   |\!\braket{\chi_1|\bm{u}^{\tiny \bm{Z}}_r}\!|^2\epsilon^2)$.

\subsection{Required number of quantum SVDs for singular value and vector collection}

The singular value estimation and the two-state and three-state SWAP tests
require quantum states encoding a singular value or a singular vector.
Furthermore, it is necessary to repeat the SWAP test for each element
a certain number of times, as discussed in the previous subsection.
A single quantum SVD process and measuring the fifth register provides
a triple of a singular value and its associated left and right singular
vector states at random. Thus, in this subsection, let us consider
the number of quantum SVDs necessary for collecting $m$ each of
all dominant singular value and singular vector states.

Brayton \cite{Brayton1963} investigated the asymptotic behavior of the expected
number of trials necessary to collect $m$ copies of a set of $n$ objects
by random selection, denoted by $E_m(n)$. The probability of selecting
the $i$-th object is assumed to be given by
\begin{equation}
p_i = \int_{(i-1)/n}^{i/n} f(s) \mathrm{d}s,
\end{equation}
where the function $f$ satisfies the following conditions:
\begin{enumerate}
\item $\int_0^1 f(s) \mathrm{d}s = 1$.
\item $\min f(s) = \delta >0;\ \max f(s) < \infty$.
\item $f(s)$ is a function of bounded variation on $[0, 1]$.
\end{enumerate}
Brayton proved that if $f(s)=\delta$ on a set of intervals of length
$l>0$, the asymptotic behavior of $E_m(n)$ as $n\to\infty$ is given by
\begin{equation}
E_m(n) = \frac{n}{\delta}\left[
  \log \Gamma n + (m-1)\log\log \Gamma n + \gamma +
  (m-1)\log\frac{1}{\delta} + o(1)
\right],
\end{equation}
where $\gamma$ is the Euler--Mascheroni constant and
\begin{equation}
\Gamma = l\frac{\delta^{m-1}}{(m-1)!}.
\end{equation}

Let us define $f$ by a piecewise constant function such that
$f(s) = np_i$ for $(i-1)/n \le s < i/n$. Additionally, we denote
the minimum probability of random selection by $p_\mathrm{min}$.
In that case, $\delta = np_\mathrm{min} \le 1$ and $\Gamma \le l \le 1$.
Therefore, the expected number of necessary trials roughly scales as
\begin{equation}
E_m(n) =
O\left(
    \frac{1}{p_{\mathrm{min}}}\left[
    \log n + (m-1)\log\frac{\log n}{np_\mathrm{min}}
    \right]
\right).
\end{equation}
We can further rewrite this asymptotic equation using the maximum probability
of random selection, $p_\mathrm{max}$, as
\begin{equation}
E_m(n) =
O\left(
    \frac{p_{\mathrm{max}}}{p_{\mathrm{min}}}\left[
    n\log n + n(m-1)\log\left(\frac{p_\mathrm{max}}{p_\mathrm{min}}\log n\right)
    \right]
\right),
\end{equation}
because $1 \le np_\mathrm{max}$ holds.

In the case of the singular value and singular vector collection
for matrix $\bm{Z}$, the number of objects $n$ equals $R$,
and the ratio of the maximum and minimum probabilities is given by
\begin{equation}
\frac{p_{\mathrm{max}}}{p_{\mathrm{min}}} =
\frac{\max_r (\hat{\sigma}^{\tiny \bm{Z}}_r)^2}
     {\min_r (\hat{\sigma}^{\tiny \bm{Z}}_r)^2} =
\left(\frac{\max_r \hat{\sigma}^{\tiny \bm{Z}}_r}
           {\min_r \hat{\sigma}^{\tiny \bm{Z}}_r}\right)^2.
\end{equation}
Let us define a parameter $\kappa_{\tiny \bm{Z}}$ as
\begin{equation}
\kappa_{\tiny \bm{Z}} \coloneqq
\frac{\max_r \sigma^{\tiny \bm{Z}}_r}
     {\min_r \sigma^{\tiny \bm{Z}}_r} =
\frac{\max_r \hat{\sigma}^{\tiny \bm{Z}}_r}
     {\min_r \hat{\sigma}^{\tiny \bm{Z}}_r}.
\label{eq:kappa-def}
\end{equation}
The parameter $\kappa_{\tiny \bm{Z}}$ is also known as the condition number
of $\bm{Z}$ (with rank-$R$ truncation). Using this parameter, we can write
the required number of quantum SVDs for collecting $m$ each of all dominant
singular value and singular vector states of $\bm{Z}$ as
\begin{equation}
E_m(R) =
O\left(
    \kappa_{\tiny \bm{Z}}^2 \left[
    R\log R + (m-1)R\log\left(\kappa_{\tiny \bm{Z}}^2\log R\right)
    \right]
\right).
\label{eq:Brayton}
\end{equation}

Note that since the qDMD algorithm relies on the rank-$R$ approximation of
$\hat{\bm{Z}}$, the probability of selecting a singular value other than
the $R$ dominant singular values may be non-zero. However, as we determine $R$
such that the error of the rank-$R$ approximation is less than a specified
tolerance in terms of the Frobenius norm, the probability of selecting one of
$R$ dominant singular values, given by
$\sum_{r=1}^R (\hat{\sigma}^{\tiny \bm{Z}}_r)^2$, is $O(1)$.
Thus we can neglect the probability of selecting other minor singular values.

\subsection{Required number of quantum SVDs for estimating $\tilde{\bm{K}}^\prime$}

Now, we evaluate the total number of quantum SVDs necessary for estimating
the whole matrix $\tilde{\bm{K}}^\prime$. We first note that the factor
$\|\bm{X}^\prime\|_\mathrm{F}/\|\bm{X}\|_\mathrm{F}$ and the normalized
singular value matrices $\hat{\bm{\Sigma}}$ and $\hat{\bm{\Sigma}}^\prime$
can be estimated in the course of preparing singular vector states for
the SWAP tests. In other words, we can include the costs of these estimations
into those of preparing singular vector states. Therefore, we only consider
the computational costs for the inner product estimation.

According to Table~\ref{tbl:estimation-processes}, the required number of repetitions
of the three-state SWAP test depends on inner products between singular
vector states and reference states. For the simplicity of the notation,
let us introduce a parameter $\zeta$ defined as
\begin{align}
\zeta &\coloneqq \min\{\zeta_1, \zeta_2\}, \label{eq:zeta-def} \\
\zeta_1 &\coloneqq
\min \left\{|\!\braket{\chi_1|\bm{u}^{\tiny \bm{Z}}_r}\!|^2\ \middle|\
            \bm{Z}\in\{\bm{X}, \bm{X}^\prime, [\bm{X}\ \bm{X}^\prime]\},\
            r \in \{1,\dots, R\} \right\}, \label{eq:zeta1-def} \\
\zeta_2 &\coloneqq
\min \left\{|\!\braket{\chi_2|\bm{v}^{\tiny \bm{Z} *}_r}\!|^2\ \middle|\
            \bm{Z}\in\{\bm{X}, \bm{X}^\prime\},\
            r \in \{1,\dots, R\} \right\}.
\end{align}
Then, the required number of repetitions of the three-state SWAP test scales
as $O(1/(\zeta \epsilon)^2)$. In addition, each singular vector is involved in
$O(R)$ different inner products to be estimated by the three-state SWAP test.
Consequently, the three-state SWAP test requires $O(R/(\zeta \epsilon)^2)$
copies of each singular vector state in total. Furthermore, the two-state
SWAP test requires $O(1/\epsilon^2)$ copies of each left singular vector
state. Because $R/\zeta^2 \ge 1$, the total copy number of each singular
vector state necessary for the two-state and three-state SWAP tests is
roughly estimated as $O(R/(\zeta \epsilon)^2)$. Due to Eq.~\eqref{eq:Brayton},
we get the order estimate of the total number of quantum SVDs necessary
for the SWAP tests in the qDMD algorithm:
\begin{equation}
O\left(
    \left(\frac{\kappa R}{\zeta \epsilon}\right)^2
    \log\left(\kappa^2\log R\right)
\right), \label{eq:num_qSVD_1}
\end{equation}
where we define $\kappa$ by $\max_{\tiny \bm{Z}} \kappa_{\tiny \bm{Z}}$
for brevity.

According to Table~\ref{tbl:estimation-processes}, the total number of repetitions
of the quantum process for estimating $2R$ inner products
$\{\braket{\chi_2|\bm{v}^{\tiny \bm{Z}*}_r}\}$ is
\begin{equation}
O\left(\frac{\eta}{\zeta_1} \left(\frac{\kappa R}{\epsilon}\right)^2 \right),
\label{eq:num_qSVD_2}
\end{equation}
where we define parameter $\eta$ as
\begin{equation}
\eta \coloneqq
\frac{1}{\min\{P_{\tiny \bm{X}}, P_{\tiny \bm{X}^\prime}\}},
\end{equation}
and we use the equation
\begin{equation}
\frac{1}{\min_r (\hat{\sigma}^{\tiny \bm{Z}}_r)^2}
= O(R \kappa_{\tiny \bm{Z}}^2).
\label{eq:kappa-sigma_min-relation}
\end{equation}
Eq.~\eqref{eq:kappa-sigma_min-relation} is equivalent to the equation
$1/p_\mathrm{min} \le R (p_\mathrm{max}/p_\mathrm{min})$,
which is used to derive Eq.~\eqref{eq:Brayton} in the previous subsection.
Since the process contains one quantum SVD process, this total number of
repetitions given by Eq.~\eqref{eq:num_qSVD_2} equals the total number of
quantum SVDs for estimating $\{\braket{\chi_2|\bm{v}^{\tiny \bm{Z}*}_r}\}$.
Additionally, we note that $\eta$ is $O(1)$ due to the following reason:
As stated in the main text, it is assumed that $\Delta t \le 1/\|\bm{A}\|_2$.
Hence, the norm change of a state vector due to the one-step time evolution
is bounded as follows:
\begin{equation}
\|\bm{x}_j^\prime\|_2
\approx \left\| \mathrm{e}^{\Delta t \bm{A}} \bm{x}_j \right\|_2
\le \mathrm{e}^{\Delta t \|\bm{A}\|_2} \|\bm{x}_j\|_2
\le \mathrm{e} \|\bm{x}_j\|_2,
\end{equation}
and
\begin{equation}
\|\bm{x}_j\|_2
\approx \left\| \mathrm{e}^{-\Delta t \bm{A}} \bm{x}_j^\prime \right\|_2
\le \mathrm{e}^{\Delta t \|\bm{A}\|_2} \|\bm{x}_j^\prime\|_2
\le \mathrm{e} \|\bm{x}_j^\prime\|_2.
\end{equation}
Since $\|\bm{X}\|_\mathrm{F}=\sqrt{\sum_{j=0}^{M-1} \|\bm{x}_j\|_2^2}$ and
$\|\bm{X}^\prime\|_\mathrm{F}=\sqrt{\sum_{j=0}^{M-1} \|\bm{x}_j^\prime\|_2^2}$,
we obtain the inequality
\begin{equation}
\frac{1}{\mathrm{e}} \|\bm{X}\|_\mathrm{F}
\lessapprox \|\bm{X}^\prime\|_\mathrm{F}
\lessapprox \mathrm{e} \|\bm{X}\|_\mathrm{F}.
\end{equation}
This implies that $P_{\tiny \bm{X}}$ and $P_{\tiny \bm{X}^\prime}$ are
$\Omega(1)$, where the $\Omega$-symbol signifies an asymptotic lower bound.
Therefore, $\eta$ is $O(1)$. Due to the facts that $\eta = O(1)$ and
that $\zeta \le \zeta_1 < 1$, the cost estimate in Eq.~\eqref{eq:num_qSVD_2} is
less than that in Eq.~\eqref{eq:num_qSVD_1}.

In summary, the number of quantum SVDs necessary for estimating
$\tilde{\bm{K}}^\prime$ with precision $\epsilon$ is
\begin{equation}
O\left(
    \left(\frac{\kappa R}{\zeta \epsilon}\right)^2
    \log\left(\kappa^2\log R\right)
\right).
\label{eq:num_qSVD}
\end{equation}
Because one quantum SVD process requires $O(1/\epsilon^2)$ copies of
a data matrix \cite{Kimmel2017}, the required number of differential equation
solutions is
\begin{equation}
O\left(
    \frac{1}{\epsilon^4}\left(\frac{\kappa R}{\zeta}\right)^2
    \log\left(\kappa^2\log R\right)
\right).
\end{equation}
The gate complexity is larger than the required number of differential equation
solutions by a factor of $T \operatorname{poly} \log(NM/\epsilon)$.

\section{Reference State Preparation for the Inner Product Estimation}

One may use any reference states $\ket{\chi_1}$ and $\ket{\chi_2}$, provided
that $\braket{\chi_1|\bm{u}} \neq 0$ and $\braket{\chi_2|\bm{v}^*} \neq 0$
for all relevant left and right singular vectors $\bm{u}$ and $\bm{v}$.
This condition is the necessary and sufficient condition so that
the inner product estimation through the three-state SWAP test is possible.
Regardless of the choice of $\ket{\chi_1}$ and $\ket{\chi_2}$, estimates of
$\tilde{\bm{K}}^\prime$ are identical within a tolerant error.
However, the choice of $\ket{\chi_1}$ and $\ket{\chi_2}$ affects
the algorithm's efficiency: the computational complexity of estimating
$\tilde{\bm{K}}^\prime$ is influenced by the choice of $\ket{\chi_1}$
and $\ket{\chi_2}$ through the parameter $\zeta$, as discussed in
the previous section; the larger $\zeta$ is, the smaller the required
number of quantum SVDs is. Moreover, if the reference state preparation has
a runtime greater than $O(\operatorname{poly}\log N)$, the qDMD algorithm
cannot achieve an exponential speedup with respect to $N$ in total.

In this section, we present an example of quantum procedures preparing
$\ket{\chi_1}$ and $\ket{\chi_2}$. The quantum circuits of these procedures
can be constructed and executed in $O(\operatorname{poly}\log N)$ time,
ensuring that $1/\zeta = O(\operatorname{poly}R)$. Therefore, this example
theoretically guarantees that the qDMD algorithm achieves an exponential
speedup with respect to $N$ in estimating $\tilde{\bm{K}}^\prime$.
Additionally, we remark on other possible methods of the reference state
preparation at the end of this section.

\subsection{Choice of $\ket{\chi_1}$ and $\ket{\chi_2}$: An example}

We define $\ket{\chi_1}$ such that $1/\zeta_1 = O(\operatorname{poly}R)$.
Specifically, it is designed here to satisfy the following conditions:
\begin{align}
|\!\braket{\chi_1|\bm{u}^{\tiny [\bm{X}\ \bm{X}^\prime]}_r}\!|^2
& = \Omega\left(\frac{1}{R}\right), \\
|\!\braket{\chi_1|\bm{u}^{\tiny \bm{X}}_r}\!|^2
& = \Omega\left(\frac{1}{R^2}\right), \\
|\!\braket{\chi_1|\bm{u}^{\tiny \bm{X}^\prime}_r}\!|^2
& = \Omega\left(\frac{1}{R^3}\right),
\end{align}
for any $r\in\{1,2,\dots,R\}$. Here, the $\Omega$-symbol signifies
an asymptotic lower bound.

Initially, we construct a quantum state $\ket{\chi_1^{(0)}}$ such that
$|\!\braket{\chi_1^{(0)}|\bm{u}^{\tiny [\bm{X}\ \bm{X}^\prime]}_r}\!|^2
 = \Omega(1/R)$ for any $r\in\{1,2,\dots,R\}$. The superposition of
all left singular vector states of $[\bm{X}\ \bm{X}^\prime]$ with
equal weights satisfies this condition. Thus we define $\ket{\chi_1^{(0)}}$ as
\begin{equation}
\ket{\chi_1^{(0)}} \coloneqq
\frac{1}{\sqrt{R}}
\sum_{r=1}^R \ket{\bm{u}^{\tiny [\bm{X}\ \bm{X}^\prime]}_r}.
\end{equation}

Next, we modify $\ket{\chi_1^{(0)}}$ to a quantum state $\ket{\chi_1^{(1)}}$
that satisfies
$|\!\braket{\chi_1^{(1)}|\bm{u}^{\tiny \bm{X}}_r}\!|^2 = \Omega(1/R^2)$
as well as
$|\!\braket{\chi_1^{(1)}|\bm{u}^{\tiny [\bm{X}\ \bm{X}^\prime]}_r}\!|^2
 = \Omega(1/R)$ for any $r\in\{1,2,\dots,R\}$. Let $S_1^{(1)}$ be a set
of left singular vector states of $\bm{X}$ defined by
\begin{equation}
S_1^{(1)} \coloneqq
\left\{
     \ket{\bm{u}^{\tiny \bm{X}}_r}
     \middle| r \in \{1,2,\dots,R\},
     |\!\braket{\chi_1^{(0)}|\bm{u}^{\tiny \bm{X}}_r}\!| \le \frac{1}{4R}
\right\}.
\end{equation}
This set is a collection of $\ket{\bm{u}^{\tiny \bm{X}}_r}$ whose
overlap with the temporary reference state $\ket{\chi_1^{(0)}}$, i.e.,
$|\!\braket{\chi_1^{(0)}|\bm{u}^{\tiny \bm{X}}_r}\!|^2$, is small compared
with $1/R^2$. To ensure that $\ket{\bm{u}^{\tiny \bm{X}}_r} \in S_1^{(1)}$ has
$\Omega(1/R^2)$ overlap with the reference state, we add a correction term
$\ket{\phi_1^{(1)}}$ to the temporary reference state $\ket{\chi_1^{(0)}}$:
\begin{equation}
\ket{\phi_1^{(1)}} \coloneqq
\frac{1}{\sqrt{|S_1^{(1)}|}}
\sum_{\ket{\bm{u}^{\tiny \bm{X}}_r} \in S_1^{(1)}}
\ket{\bm{u}^{\tiny \bm{X}}_r},
\end{equation}
and
\begin{equation}
\ket{\chi_1^{(1)}} \coloneqq
\mathcal{C}_1^{(1)}\left[
     \ket{\chi_1^{(0)}} + \frac{1}{2\sqrt{R}}\ket{\phi_1^{(1)}}
\right].
\end{equation}
Here, $\mathcal{C}_1^{(1)}$ is the normalizing constant. The coefficient
$1/2\sqrt{R}$ of $\ket{\phi_1^{(1)}}$ is introduced to ensure that
this correction is small enough not to violate the condition
$|\!\braket{\chi_1^{(1)}|\bm{u}^{\tiny [\bm{X}\ \bm{X}^\prime]}_r}\!|^2
 = \Omega(1/R)$. We can confirm that $\ket{\chi_1^{(1)}}$ satisfies
the aforementioned conditions on inner products as follows:
The normalizing constant $\mathcal{C}_1^{(1)}$ is $\Omega(1)$ because
\begin{equation}
\begin{split}
\mathcal{C}_1^{(1)} &=
\frac{1}{\|\!\ket{\chi_1^{(0)}} + \frac{1}{2\sqrt{R}}\ket{\phi_1^{(1)}}\!\|} \\
&\ge \frac{1}{\|\!\ket{\chi_1^{(0)}}\!\| +
              \frac{1}{2\sqrt{R}}\|\!\ket{\phi_1^{(1)}}\!\|} \\
&= \frac{1}{1+\frac{1}{2\sqrt{R}}} \\
&\ge \frac{2}{3}.
\end{split}
\end{equation}
For any $\ket{\bm{u}^{\tiny [\bm{X}\ \bm{X}^\prime]}_r}$,
\begin{equation}
\begin{split}
|\!\braket{\chi_1^{(1)}|\bm{u}^{\tiny [\bm{X}\ \bm{X}^\prime]}_r}\!|
&\ge \mathcal{C}_1^{(1)}\left[
    |\!\braket{\chi_1^{(0)}|\bm{u}^{\tiny [\bm{X}\ \bm{X}^\prime]}_r}\!| -
    \frac{1}{2\sqrt{R}}
    |\!\braket{\phi_1^{(1)}|\bm{u}^{\tiny [\bm{X}\ \bm{X}^\prime]}_r}\!|
\right] \\
&\ge \frac{2}{3}\left[\frac{1}{\sqrt{R}} - \frac{1}{2\sqrt{R}}\right] \\
&= \frac{1}{3\sqrt{R}}.
\end{split}
\end{equation}
The first inequality follows from the triangle inequality.
Likewise, for any $\ket{\bm{u}^{\tiny \bm{X}}_r} \in S_1^{(1)}$,
\begin{equation}
\begin{split}
|\!\braket{\chi_1^{(1)}|\bm{u}^{\tiny \bm{X}}_r}\!|
&\ge \mathcal{C}_1^{(1)}\left[
    \frac{1}{2\sqrt{R}}|\!\braket{\phi_1^{(1)}|\bm{u}^{\tiny \bm{X}}_r}\!| -
    |\!\braket{\chi_1^{(0)}|\bm{u}^{\tiny \bm{X}}_r}\!|
\right] \\
&\ge \frac{2}{3}\left[
     \frac{1}{2\sqrt{R}}\frac{1}{\sqrt{|S_1^{(1)}|}} - \frac{1}{4R}
\right] \\
&\ge \frac{2}{3}\left[
     \frac{1}{2\sqrt{R}}\frac{1}{\sqrt{R}} - \frac{1}{4R}
\right] \\
&= \frac{1}{6R}.
\end{split}
\end{equation}
Any $\ket{\bm{u}^{\tiny \bm{X}}_r} \notin S_1^{(1)}$ is orthogonal to
$\ket{\phi_1^{(1)}} \in \operatorname{span}S_1^{(1)}$.
Consequently, for any $\ket{\bm{u}^{\tiny \bm{X}}_r} \notin S_1^{(1)}$,
\begin{equation}
|\!\braket{\chi_1^{(1)}|\bm{u}^{\tiny \bm{X}}_r}\!|
= \mathcal{C}_1^{(1)} |\!\braket{\chi_1^{(0)}|\bm{u}^{\tiny \bm{X}}_r}\!|
\ge \frac{2}{3}\frac{1}{4R} 
= \frac{1}{6R}.
\end{equation}
Therefore, $\ket{\chi_1^{(1)}}$ satisfies
$|\!\braket{\chi_1^{(1)}|\bm{u}^{\tiny [\bm{X}\ \bm{X}^\prime]}_r}\!|^2
 = \Omega(1/R)$ and
$|\!\braket{\chi_1^{(1)}|\bm{u}^{\tiny \bm{X}}_r}\!|^2 = \Omega(1/R^2)$
for any $r\in\{1,2,\dots,R\}$.

Finally, we modify $\ket{\chi_1^{(1)}}$ to a quantum state $\ket{\chi_1^{(2)}}$
that also satisfies
$|\!\braket{\chi_1^{(2)}|\bm{u}^{\tiny \bm{X}^\prime}_r}\!|^2 = \Omega(1/R^3)$.
Let us define $S_1^{(2)}$ as
\begin{equation}
S_1^{(2)} \coloneqq
\left\{
    \ket{\bm{u}^{\tiny \bm{X}^\prime}_r}
    \middle| r \in \{1,2,\dots,R\},
    |\!\braket{\chi_1^{(1)}|\bm{u}^{\tiny \bm{X}^\prime}_r}\!|
    \le \frac{1}{14R\sqrt{R}}
\right\}.
\end{equation}
Then, we define $\ket{\chi_1^{(2)}}$ by adding a correction term
$\ket{\phi_1^{(2)}}$ to $\ket{\chi_1^{(1)}}$:
\begin{equation}
\ket{\phi_1^{(2)}} \coloneqq
\frac{1}{\sqrt{|S_1^{(2)}|}}
\sum_{\ket{\bm{u}^{\tiny \bm{X}^\prime}_r} \in S_1^{(2)}}
\ket{\bm{u}^{\tiny \bm{X}^\prime}_r},
\end{equation}
and
\begin{equation}
\ket{\chi_1^{(2)}} \coloneqq
\mathcal{C}_1^{(2)}\left[
     \ket{\chi_1^{(1)}} + \frac{1}{7R}\ket{\phi_1^{(2)}}
\right],
\end{equation}
where $\mathcal{C}_1^{(2)}$ is the normalizing constant. We can confirm
that $\ket{\chi_1^{(2)}}$ satisfies the aforementioned conditions on
inner products as follows: The normalizing constant $\mathcal{C}_1^{(2)}$
is $\Omega(1)$ because
\begin{equation}
\begin{split}
\mathcal{C}_1^{(2)}
&= \frac{1}{\|\!\ket{\chi_1^{(1)}} + \frac{1}{7R}\ket{\phi_1^{(2)}}\!\|} \\
&\ge \frac{1}{\|\!\ket{\chi_1^{(1)}}\!\| +
              \frac{1}{7R}\|\!\ket{\phi_1^{(2)}}\!\|} \\
&\ge \frac{1}{1 + \frac{1}{7R}} \\
&\ge \frac{7}{8}.
\end{split}
\end{equation}
For any $\ket{\bm{u}^{\tiny [\bm{X}\ \bm{X}^\prime]}_r}$,
\begin{equation}
\begin{split}
|\!\braket{\chi_1^{(2)}|\bm{u}^{\tiny [\bm{X}\ \bm{X}^\prime]}_r}\!|
&\ge \mathcal{C}_1^{(2)}\left[
     |\!\braket{\chi_1^{(1)}|\bm{u}^{\tiny [\bm{X}\ \bm{X}^\prime]}_r}\!| -
     \frac{1}{7R}
     |\!\braket{\phi_1^{(2)}|\bm{u}^{\tiny [\bm{X}\ \bm{X}^\prime]}_r}\!|
\right] \\
&\ge \frac{7}{8}\left[\frac{1}{3\sqrt{R}} - \frac{1}{7R}\right] \\
&\ge \frac{7}{8}\left[\frac{1}{3\sqrt{R}} - \frac{1}{7\sqrt{R}}\right] \\
&= \frac{1}{6\sqrt{R}}.
\end{split}
\end{equation}
For any $\ket{\bm{u}^{\tiny \bm{X}}_r}$,
\begin{equation}
\begin{split}
|\!\braket{\chi_1^{(2)}|\bm{u}^{\tiny \bm{X}}_r}\!|
&\ge \mathcal{C}_1^{(2)}\left[
     |\!\braket{\chi_1^{(1)}|\bm{u}^{\tiny \bm{X}}_r}\!| -
     \frac{1}{7R}
     |\!\braket{\phi_1^{(2)}|\bm{u}^{\tiny \bm{X}}_r}\!|
\right] \\
&\ge \frac{7}{8}\left[\frac{1}{6R} - \frac{1}{7R}\right] \\
&= \frac{1}{48R}.
\end{split}
\end{equation}
For any $\ket{\bm{u}^{\tiny \bm{X}^\prime}_r} \in S_1^{(2)}$,
\begin{equation}
\begin{split}
|\!\braket{\chi_1^{(2)}|\bm{u}^{\tiny \bm{X}^\prime}_r}\!|
&\ge \mathcal{C}_1^{(2)}\left[
    \frac{1}{7R}
    |\!\braket{\phi_1^{(2)}|\bm{u}^{\tiny \bm{X}^\prime}_r}\!| -
    |\!\braket{\chi_1^{(1)}|\bm{u}^{\tiny \bm{X}^\prime}_r}\!|
\right] \\
&\ge \frac{7}{8}\left[
     \frac{1}{7R}\frac{1}{\sqrt{|S_1^{(2)}|}} - \frac{1}{14R\sqrt{R}}
\right] \\
&\ge \frac{7}{8}\left[
     \frac{1}{7R}\frac{1}{\sqrt{R}} - \frac{1}{14R\sqrt{R}}
\right] \\
&= \frac{1}{16R\sqrt{R}}.
\end{split}
\end{equation}
For any $\ket{\bm{u}^{\tiny \bm{X}^\prime}_r} \notin S_1^{(2)}$,
\begin{equation}
|\!\braket{\chi_1^{(2)}|\bm{u}^{\tiny \bm{X}^\prime}_r}\!|
= \mathcal{C}_1^{(2)}
  |\!\braket{\chi_1^{(1)}|\bm{u}^{\tiny \bm{X}^\prime}_r}\!|
\ge \frac{7}{8}\frac{1}{14R\sqrt{R}}
= \frac{1}{16R\sqrt{R}}.
\end{equation}
Therefore, $\ket{\chi_1^{(2)}}$ defined here can be employed as
$\ket{\chi_1}$ that satisfies $1/\zeta_1 = O(\operatorname{poly}R)$.

Likewise, a reference state $\ket{\chi_2}$ that satisfies
$1/\zeta_2 = O(\operatorname{poly}R)$ can be defined as
\begin{equation}
\ket{\chi_2} \coloneqq
\mathcal{C}_2^{(1)}\left[
     \ket{\chi_2^{(0)}} + \frac{1}{2\sqrt{R}}\ket{\phi_2^{(1)}}
\right],
\end{equation}
where $\mathcal{C}_2^{(1)}$ is the normalizing constant,
\begin{align}
\ket{\chi_2^{(0)}} &\coloneqq
\frac{1}{\sqrt{R}}\sum_{r=1}^R \ket{\bm{v}^{\tiny \bm{X} *}_r}, \\
\ket{\phi_2^{(1)}} &\coloneqq
\frac{1}{\sqrt{|S_2^{(1)}|}}
\sum_{\ket{\bm{v}^{\tiny \bm{X}^\prime *}_r} \in S_2^{(1)}}
\ket{\bm{v}^{\tiny \bm{X}^\prime *}_r},
\end{align}
and
\begin{equation}
S_2^{(1)} \coloneqq
\left\{
     \ket{\bm{v}^{\tiny \bm{X}^\prime *}_r}
     \middle| r \in \{1,2,\dots,R\},
     |\!\braket{\chi_2^{(0)}|\bm{v}^{\tiny \bm{X}^\prime *}_r}\!|
     \le \frac{1}{4R}
\right\}.
\end{equation}
Note that the qDMD algorithm does not need to estimate
$\braket{\chi_2|\bm{v}^{\tiny [\bm{X}\ \bm{X}^\prime]*}_r}$.
Consequently, the above definition of $\ket{\chi_2}$ does not contain
$\ket{\bm{v}^{\tiny [\bm{X}\ \bm{X}^\prime]*}_r}$.

\subsection{Computation of $\ket{\chi_1}$ and $\ket{\chi_2}$: An example}

We present quantum circuits that generate $\ket{\chi_1}$ and $\ket{\chi_2}$
defined in the previous subsection.

Initially, we perform pure-state tomography \cite{Verdeil2023} to estimate
all right singular vector states of $\bm{X}$, $\bm{X}^\prime$,
and $[\bm{X}\ \bm{X}^\prime]$. The pure-state tomography requires
an $O(M)$ runtime for each right singular vector state of $M$ dimensions.
Subsequently, for each right singular vector state $\ket{\bm{v}^*}$,
we construct a quantum circuit $U_{\bm{v}^*}$ such that
$U_{\bm{v}^*}\ket{0} = \ket{\bm{v}^*}$ \footnote{
  Because the pure-state tomography can not determine the global phase
  of a right singular vector state, the unitary gate $U_{\bm{v}^*}$ may act
  as $U_{\bm{v}^*}\ket{0} = \exp(-\mathrm{i}\varphi)\ket{\bm{v}^*}$ where
  $\varphi$ signifies the unknown global phase. This unknown phase factor
  will change the relative phase of the associated left and right singular
  vector states in the computed reference states. However, such phase factor
  does not violate the condition $1/\zeta = O(\operatorname{poly} R)$ proven
  in the previous subsection. Thus, we omit the phase factor for simplicity
  in this section.
}. When the classical data of
the amplitudes of $\ket{\bm{v}^*}$ is given, such a quantum circuit can be
implemented with a gate complexity of $O(\operatorname{poly}\log M)$
\cite{Kerenidis2017,Mitarai2019}.

Next, we compute $\ket{\chi_1^{(0)}}$ with the following steps:
(1) Prepare $\ket{\mathrm{SVD}(\hat{[\bm{X}\ \bm{X}^\prime]})}$.
(2) Uncompute the second to fourth registers of
$\ket{\mathrm{SVD}(\hat{[\bm{X}\ \bm{X}^\prime]})}$ by
controlled-$U_{\bm{v}^{\tiny [\bm{X}\ \bm{X}^\prime]*}_r}^\dagger$ operations:
\begin{equation}
\begin{split}
\ket{\mathrm{SVD}(\hat{[\bm{X}\ \bm{X}^\prime]})}
&= \sum_{r=1}^R \hat{\sigma}^{\tiny [\bm{X}\ \bm{X}^\prime]}_r
   \ket{\bm{u}^{\tiny [\bm{X}\ \bm{X}^\prime]}_r}_1\!
   \ket{\bm{v}^{\tiny [\bm{X}\ \bm{X}^\prime] *}_r}_{234}\!
   \ket{(\hat{\sigma}^{\tiny [\bm{X}\ \bm{X}^\prime]}_r)^2}_5 \\
&\mapsto \sum_{r=1}^R \hat{\sigma}^{\tiny [\bm{X}\ \bm{X}^\prime]}_r
         \ket{\bm{u}^{\tiny [\bm{X}\ \bm{X}^\prime]}_r}_1\!
         U_{\bm{v}^{\tiny [\bm{X}\ \bm{X}^\prime] *}_r}^\dagger
         \ket{\bm{v}^{\tiny [\bm{X}\ \bm{X}^\prime] *}_r}_{234}\!
         \ket{(\hat{\sigma}^{\tiny [\bm{X}\ \bm{X}^\prime]}_r)^2}_5 \\
&= \sum_{r=1}^R \hat{\sigma}^{\tiny [\bm{X}\ \bm{X}^\prime]}_r
   \ket{\bm{u}^{\tiny [\bm{X}\ \bm{X}^\prime]}_r}_1\!
   \ket{0}_{234}\!
   \ket{(\hat{\sigma}^{\tiny [\bm{X}\ \bm{X}^\prime]}_r)^2}_5.
\end{split}
\end{equation}
Here, each $U_{\bm{v}^{\tiny [\bm{X}\ \bm{X}^\prime]*}_r}^\dagger$ is
applied to the second to fourth registers of
$\ket{\mathrm{SVD}(\hat{[\bm{X}\ \bm{X}^\prime]})}$ conditionally
on the fifth register.
(3) Add an ancilla qubit (the sixth register).
(4) Compute the following state by controlled rotations of the ancilla
qubit conditionally on the fifth register:
\begin{equation}
\sum_{r=1}^R \hat{\sigma}^{\tiny [\bm{X}\ \bm{X}^\prime]}_r
\ket{\bm{u}^{\tiny [\bm{X}\ \bm{X}^\prime]}_r}_1\!
\ket{0}_{234}\!
\ket{(\hat{\sigma}^{\tiny [\bm{X}\ \bm{X}^\prime]}_r)^2}_5\!
\left[
     \frac{a_0}{\hat{\sigma}^{\tiny [\bm{X}\ \bm{X}^\prime]}_r\sqrt{R}}
     \ket{0}_6 +
     \sqrt{1-\left|
          \frac{a_0}{\hat{\sigma}^{\tiny [\bm{X}\ \bm{X}^\prime]}_r\sqrt{R}}
          \right|^2}
     \ket{1}_6
\right],
\end{equation}
where $a_0$ is a constant such that
\begin{equation}
\frac{a_0}{\hat{\sigma}^{\tiny [\bm{X}\ \bm{X}^\prime]}_r\sqrt{R}} \le 1,
\quad \forall r \in \{1,2,\dots,R\}.  
\end{equation}
This condition is the necessary and sufficient condition for
the controlled rotations to be possible and is equivalent to
\begin{equation}
a_0 \le \sqrt{R} \min_r\hat{\sigma}^{\tiny [\bm{X}\ \bm{X}^\prime]}_r.
\end{equation}
(5) Uncompute the fifth register by the inverse transform of
the quantum SVD process. Specifically, the inverse process consists of
the inverse quantum Fourier transform of the fifth register and
the density matrix exponentiation multiplying
$\exp(-\mathrm{i}\operatorname{tr}_{234}
      (\ket{\hat{[\bm{X}\ \bm{X}^\prime]}}\!
       \bra{\hat{[\bm{X}\ \bm{X}^\prime]}})t)$
to the first register conditioned on the fifth register. Here,
$\operatorname{tr}_{234}$ denotes the partial trace with respect to
the second to fourth registers. See \cite{Schuld2016} for the details of
the quantum SVD process.
(6) Measure the second to sixth registers. If all measured values are zero,
we obtain $\ket{\chi_1^{(0)}}_1\!\ket{0}_{2:6}$.
The success probability is $a_0^2$.
When $\sqrt{R}\min_r \hat{\sigma}^{\tiny [\bm{X}\ \bm{X}^\prime]}_r$ is
employed as $a_0$, an asymptotic lower bound of the success probability
can be evaluated as
\begin{equation}
a_0^2
= R (\min_r \hat{\sigma}^{\tiny [\bm{X}\ \bm{X}^\prime]}_r)^2
= \Omega\left(R\frac{1}{R\kappa_{\tiny [\bm{X}\ \bm{X}^\prime]}^2}\right)
= \Omega\left(\frac{1}{\kappa^2}\right).
\end{equation}
Here, $\kappa_{\tiny \bm{Z}}$ denotes the condition number of $\bm{Z}$
($\bm{Z}=\bm{X}, \bm{X}^\prime, [\bm{X}\ \bm{X}^\prime]$) defined in
Eq.~\eqref{eq:kappa-def}, $\kappa\coloneqq \max_{\bm{Z}} \kappa_{\tiny \bm{Z}}$,
and we use Eq.~\eqref{eq:kappa-sigma_min-relation} to derive the lower bound.
In summary, we can compute $\ket{\chi_1^{(0)}}$ by $O(\kappa^2R)$
controlled-$U_{\bm{v}^*}^\dagger$ operations and $O(\kappa^2)$
forward and inverse quantum SVD operations.

Having established the computation for $\ket{\chi_1^{(0)}}$,
we next address the computation for $\ket{\chi_1^{(1)}}$:
(1) Determine whether each $\ket{\bm{u}^{\tiny \bm{X}}_r}$
belongs to $S_1^{(1)}$ using the two-state SWAP test between
$\ket{\chi_1^{(0)}}$ and $\ket{\bm{u}^{\tiny \bm{X}}_r}$.
If $|S_1^{(1)}| = 0$, then $\ket{\chi_1^{(1)}} = \ket{\chi_1^{(0)}}$.
Thus we consider the case that $|S_1^{(1)}| \ge 1$ below.
(2) Prepare the following state:
\begin{equation}
\ket{\hat{[\bm{X}\ \bm{X}^\prime]}}_{1:4}\ket{0}_{5}\!\ket{0}_6\!
\left[\frac{1}{\sqrt{2}}\left(\ket{0}_7 + \ket{1}_7\right)\right],
\end{equation}
where the fifth register is an ancilla register for indicating singular
values, and the sixth and seventh registers are one-qubit ancilla registers.
Note that $\ket{\hat{[\bm{X}\ \bm{X}^\prime]}}_{1:4}$ can be written as
\begin{equation}
\ket{\hat{[\bm{X}\ \bm{X}^\prime]}}_{1:4} =
\sqrt{P_{\tiny \bm{X}}}\ket{\hat{\bm{X}}}_{123}\!\ket{0}_4 +
\sqrt{P_{\tiny \bm{X}^\prime}}\ket{\hat{\bm{X}^\prime}}_{123}\!\ket{1}_4,
\end{equation}
with
\begin{equation}
P_{\tiny\bm{Z}} = \frac{\|\bm{Z}\|_\mathrm{F}^2}
                       {\|[\bm{X}\ \bm{X}^\prime]\|_\mathrm{F}^2},
\quad \bm{Z} \in \{\bm{X}, \bm{X}^\prime\}.
\end{equation}
Thus the prepared state is
\begin{equation}
\begin{split}
&\frac{1}{\sqrt{2}}\ket{\hat{[\bm{X}\ \bm{X}^\prime]}}_{1:4}\!
 \ket{0}_{5}\!\ket{0}_6\!\ket{0}_7 \\ +
&\frac{1}{\sqrt{2}}\left[
     \sqrt{P_{\tiny \bm{X}}}\ket{\hat{\bm{X}}}_{123}\!\ket{0}_4
     \ket{0}_{5}\!\ket{0}_6\!\ket{1}_7 +
     \sqrt{P_{\tiny \bm{X}^\prime}}\ket{\hat{\bm{X}^\prime}}_{123}\!\ket{1}_4
     \ket{0}_{5}\!\ket{0}_6\!\ket{1}_7
\right].
\end{split}
\end{equation}
(3) Perform the steps 1--5 of computing $\ket{\chi_1^{(0)}}$ described above
conditionally on the seventh register to get
\begin{equation}
\frac{1}{\sqrt{2}}a_0\ket{\chi_1^{(0)}}_1\!\ket{0}_{2:6}\!\ket{0}_7 +
\frac{1}{\sqrt{2}}
     \sqrt{P_{\tiny \bm{X}}}\ket{\hat{\bm{X}}}_{123}\!\ket{0}_4
     \ket{0}_{5}\!\ket{0}_6\!\ket{1}_7 + \cdots.
\end{equation}
In the computation for $\ket{\chi_1^{(1)}}$, the constant $a_0$ is
specified as
\begin{equation}
a_0 = \min\left\{
     \sqrt{R} \min_r\hat{\sigma}^{\tiny [\bm{X}\ \bm{X}^\prime]}_r,
     2\sqrt{P_{\tiny \bm{X}}R|S_1^{(1)}|}
     \min_{r \in \operatorname{ind} S_1^{(1)}}
     \hat{\sigma}^{\tiny \bm{X}}_r
\right\},
\end{equation}
where
\begin{equation}
\operatorname{ind} S_1^{(1)} \coloneqq
\left\{
     r\in\{1,2,\dots,R\} \middle| \ket{\bm{u}^{\tiny \bm{X}}_r} \in S_1^{(1)}
\right\}.
\end{equation}
The reason of this specific choice is clarified in the next step.
(4) In a similar way to computing $\ket{\chi_1^{(0)}}$,
compute $\ket{\phi_1^{(1)}}$ conditionally on the fourth and seventh
register:
\begin{equation}
\frac{1}{\sqrt{2}}a_0\ket{\chi_1^{(0)}}_1\!\ket{0}_{2:6}\!\ket{0}_7 +
\frac{1}{\sqrt{2}}\sqrt{P_{\tiny \bm{X}}}a_1\ket{\phi_1^{(1)}}_1\!
\ket{0}_{2:6}\!\ket{1}_7 + \cdots.
\end{equation}
Here, $a_1$ is a constant related to controlled-rotations and
needs to satisfy
\begin{equation}
a_1 \le \sqrt{|S_1^{(1)}|}\min_{r \in \operatorname{ind} S_1^{(1)}}
        \hat{\sigma}^{\tiny \bm{X}}_r.
\end{equation}
In the present computation, we choose $a_1$ as
\begin{equation}
a_1 = \frac{a_0}{2\sqrt{P_{\tiny \bm{X}}R}}.
\end{equation}
This choice is possible owing to the aforementioned choice of $a_0$.
Consequently, we have the following state:
\begin{equation}
\frac{1}{\sqrt{2}}a_0
\left[
     \ket{\chi_1^{(0)}}_1\!\ket{0}_{2:6}\!\ket{0}_7 +
     \frac{1}{2\sqrt{R}}\ket{\phi_1^{(1)}}_1\!\ket{0}_{2:6}\!\ket{1}_7
\right] + \cdots.
\end{equation}
(5) Apply the Hadamard gate to the seventh resister to get
\begin{equation}
\frac{1}{2}a_0
\left[
     \ket{\chi_1^{(0)}}_1\! +
     \frac{1}{2\sqrt{R}}\ket{\phi_1^{(1)}}_1
\right]\ket{0}_{2:7} + \cdots.
\end{equation}
(6) Measure the second to seventh registers. If all measured values are
zero, we obtain $\ket{\chi_1^{(1)}}_1\ket{0}_{2:7}$. The success probability
is $(a_0/2\mathcal{C}_1^{(1)})^2$. A lower bound of this probability can be
evaluated as
\begin{equation}
\begin{split}
\left(\frac{a_0}{2\mathcal{C}_1^{(1)}}\right)^2
&\ge \frac{P_{\tiny \bm{X}} R
          \min_{\bm{Z}} \min_r (\hat{\sigma}^{\tiny \bm{Z}}_r)^2}{4}
    \left[\|\!\ket{\chi_1^{(0)}}\!\| -
          \frac{1}{2\sqrt{R}}\|\!\ket{\phi_1^{(1)}}\!\|\right]^2 \\
&= \Omega\left(\frac{1}{\eta\kappa^2}\right).
\end{split}
\end{equation}
Therefore, we can compute $\ket{\chi_1^{(1)}}$ by $O(\eta\kappa^2R)$
controlled-$U_{\bm{v}^*}^\dagger$ operations and $O(\eta\kappa^2)$
forward and inverse quantum SVD operations.

The reference state $\ket{\chi_1}$ now can be computed as follows:
(1) In a similar way of computing $\ket{\chi_1^{(1)}}$, compute
\begin{equation}
\begin{split}
&\frac{1}{\sqrt{2}}a_0\ket{\chi_1^{(0)}}_1\!
\ket{0}_{23}\!\ket{0}_4\!\ket{0}_{5}\!\ket{0}_6\!\ket{0}_7 \\ +
&\frac{1}{\sqrt{2}}\sqrt{P_{\tiny \bm{X}}}a_1\ket{\phi_1^{(1)}}_1\!
 \ket{0}_{23}\!\ket{0}_4\!\ket{0}_{5}\!\ket{0}_6\!\ket{1}_7 \\ +
&\frac{1}{\sqrt{2}}\sqrt{P_{\tiny \bm{X}^\prime}}a_2\ket{\phi_1^{(2)}}_1\!
 \ket{0}_{23}\!\ket{1}_4\!\ket{0}_{5}\!\ket{0}_6\!\ket{1}_7 + \cdots.
\end{split}
\end{equation}
Here, we define the constants $a_0$, $a_1$, and $a_2$ as
\begin{align}
a_0 &= \min\left\{
     \sqrt{R} \min_r\hat{\sigma}^{\tiny [\bm{X}\ \bm{X}^\prime]}_r,
     \sqrt{2P_{\tiny \bm{X}}R|S_1^{(1)}|}
     \min_{r \in \operatorname{ind} S_1^{(1)}}
     \hat{\sigma}^{\tiny \bm{X}}_r,
     \sqrt{\frac{(7\mathcal{C}_1^{(1)})^2
                 P_{\tiny \bm{X}^\prime}R^2|S_1^{(2)}|}{2}}
     \min_{r \in \operatorname{ind} S_1^{(2)}}
     \hat{\sigma}^{\tiny \bm{X}^\prime}_r
\right\}, \\
a_1 &= a_0\sqrt{\frac{1}{2P_{\tiny \bm{X}}R}}, \\
a_2 &= a_0\sqrt{\frac{2}
                {(7\mathcal{C}_1^{(1)})^2P_{\tiny \bm{X}^\prime}R^2}},
\end{align}
where
\begin{equation}
\operatorname{ind} S_1^{(2)} \coloneqq
\left\{
     r\in\{1,2,\dots,R\} \middle|
     \ket{\bm{u}^{\tiny \bm{X}^\prime}_r} \in S_1^{(2)}
\right\}.
\end{equation}
Note that $S_1^{(2)}$ can be determined by the two-state SWAP test
between $\ket{\chi_1^{(1)}}$ and $\ket{\bm{u}^{\tiny \bm{X}^\prime}_r}$
and that the value of $\mathcal{C}_1^{(1)}$ can be estimated through
the success probability of computing $\ket{\chi_1^{(1)}}$.
Consequently, we have the following state:
\begin{equation}
\begin{split}
&\frac{a_0}{\sqrt{2}}\ket{\chi_1^{(0)}}_1\!\ket{0}_{2:6}\!\ket{0}_7 \\ +
&a_0\left[
     \frac{1}{2\sqrt{R}}\ket{\phi_1^{(1)}}_1\!
     \ket{0}_{23}\!\ket{0}_4\!\ket{0}_{56}\!\ket{1}_7 +
     \frac{1}{7\mathcal{C}_1^{(1)}R}\ket{\phi_1^{(2)}}_1\!
     \ket{0}_{23}\!\ket{1}_4\!\ket{0}_{56}\!\ket{1}_7
\right] + \cdots.
\end{split}
\end{equation}
(2) Apply the Hadamard gate to the fourth register conditionally on
the seventh register to get
\begin{equation}
\frac{a_0}{\sqrt{2}}\left[
     \ket{\chi_1^{(0)}}_1\!\ket{0}_{2:6}\!\ket{0}_7 +
     \frac{1}{2\sqrt{R}}\ket{\phi_1^{(1)}}_1\!\ket{0}_{2:6}\!\ket{1}_7 +
     \frac{1}{7\mathcal{C}_1^{(1)}R}\ket{\phi_1^{(2)}}_1\!
     \ket{0}_{2:6}\!\ket{1}_7
\right] + \cdots.
\end{equation}
(3) Apply the Hadamard gate to the seventh register to get
\begin{equation}
\frac{a_0}{2\mathcal{C}_1^{(1)}}\left[
     \mathcal{C}_1^{(1)}\left(
          \ket{\chi_1^{(0)}}_1 + \frac{1}{2\sqrt{R}}\ket{\phi_1^{(1)}}_1
     \right)
     + \frac{1}{7R}\ket{\phi_1^{(2)}}_1
\right]\ket{0}_{2:7} + \cdots.
\end{equation}
(4) Measure the second to seventh registers. If all measured values are zero,
we obtain $\ket{\chi_1}_1\!\ket{0}_{2:7}$. The success probability is
$(a_0/\mathcal{C}_1^{(2)}\mathcal{C}_1^{(1)})^2$, which is
$\Omega(1/\eta\kappa^2)$.
Therefore, we can compute $\ket{\chi_1}$ by $O(\eta\kappa^2R)$
controlled-$U_{\bm{v}^*}^\dagger$ operations and $O(\eta\kappa^2)$
forward and inverse quantum SVD operations.

Next, let us consider how to generate $\ket{\chi_2^{(0)}}$:
(1) Prepare the following state:
\begin{equation}
\frac{1}{\sqrt{R}}
\sum_{r=1}^R\ket{0}_{23}\!\ket{(\hat{\sigma}^{\tiny \bm{X}}_r)^2}_5.
\end{equation}
(2) Apply $U_{\bm{v}^{\tiny \bm{X} *}_r}$ conditionally on
the fifth register to get:
\begin{equation}
\frac{1}{\sqrt{R}}\sum_{r=1}^R
U_{\bm{v}^{\tiny \bm{X} *}_r}\ket{0}_{23}\!
\ket{(\hat{\sigma}^{\tiny \bm{X}}_r)^2}_5 =
\frac{1}{\sqrt{R}}\sum_{r=1}^R
\ket{\bm{v}^{\tiny \bm{X} *}_r}_{23}\!
\ket{(\hat{\sigma}^{\tiny \bm{X}}_r)^2}_5.
\end{equation}
(3) Uncompute the fifth register by the inverse quantum SVD:
\begin{equation}
\frac{1}{\sqrt{R}}\sum_{r=1}^R
\ket{\bm{v}^{\tiny \bm{X} *}_r}_{23}\!\ket{0}_5.
\end{equation}
Here, the inverse quantum SVD consists of the inverse quantum Fourier
transform of the fifth register and the density matrix exponentiation
multiplying
$\exp(-\mathrm{i} \operatorname{tr}_1
       (\ket{\hat{\bm{X}}}\!\bra{\hat{\ket{\bm{X}}}})t)$
to the second and third register conditioned on the fifth register.
The symbol $\operatorname{tr}_1$ signifies the partial trace with
respect to the first register. Thus we can compute $\ket{\chi_2^{(0)}}$
by $R$ controlled-$U_{\bm{v}^*}$ operations and one inverse quantum
SVD operation.

Lastly, we present the computation of $\ket{\chi_2}$:
(1) Prepare the following state:
\begin{equation}
\sqrt{\frac{1}{1+\frac{1}{4R}}}\left[
     \frac{1}{\sqrt{R}}\sum_{r=1}^R
     \ket{0}_{23}\!\ket{0}_4\!\ket{(\hat{\sigma}^{\tiny \bm{X}}_r)^2}_5 +
     \frac{1}{2\sqrt{R|S_2^{(1)}|}}\sum_{r \in \operatorname{ind} S_2^{(1)}}
     \ket{0}_{23}\!\ket{1}_4\!\ket{(\hat{\sigma}^{\tiny \bm{X}^\prime}_r)^2}_5
\right],
\end{equation}
where
\begin{equation}
\operatorname{ind} S_2^{(1)} \coloneqq
\left\{
     r\in\{1,2,\dots,R\} \middle|
     \ket{\bm{v}^{\tiny \bm{X}^\prime *}_r} \in S_2^{(1)}
\right\}.
\end{equation}
(2) Compute $\ket{\chi_2^{(0)}}$ and $\ket{\phi_2^{(1)}}$ conditionally
on the fourth register in a similar way to the steps 2 and 3 of computing
$\ket{\chi_2^{(0)}}$ described above:
\begin{equation}
\sqrt{\frac{1}{1+\frac{1}{4R}}}\left[
     \frac{1}{\sqrt{R}}\sum_{r=1}^R
     \ket{\bm{v}^{\tiny \bm{X} *}_r}_{23}\ket{0}_4\!\ket{0}_5 +
     \frac{1}{2\sqrt{R|S_2^{(1)}|}}\sum_{r \in \operatorname{ind} S_2^{(1)}}
     \ket{\bm{v}^{\tiny \bm{X}^\prime *}_r}_{23}\!\ket{1}_4\!\ket{0}_5
\right].
\end{equation}
(3) Apply the Hadamard gate to the fourth register. Then we have
\begin{equation}
\frac{1}{\sqrt{2}}\sqrt{\frac{1}{1+\frac{1}{4R}}}\left[
     \frac{1}{\sqrt{R}}\sum_{r=1}^R
     \ket{\bm{v}^{\tiny \bm{X} *}_r}_{23} +
     \frac{1}{2\sqrt{R|S_2^{(1)}|}}\sum_{r \in \operatorname{ind} S_2^{(1)}}
     \ket{\bm{v}^{\tiny \bm{X}^\prime *}_r}_{23}
\right]\!\ket{0}_{45} + \cdots.
\end{equation}
(4) Measure the fourth and fifth registers. If the outcome corresponds to
all zeros, we obtain $\ket{\chi_2}_{23}\!\ket{0}_{45}$. The success probability
is $1/[(2+1/2R)(C_2^{(1)})^2]$, which is $\Omega(1)$. Therefore we can compute
$\ket{\chi_2}$ by $O(R)$ controlled-$U_{\bm{v}^*}$ operations and $O(1)$ inverse
quantum SVD operations.

The presented quantum processes that compute $\ket{\chi_1}$ and $\ket{\chi_2}$
are not unitary because they include post selection steps. Instead, they are
block encodings of the unitary gates $U_{\chi_1}$ and $U_{\chi_2}$.
Even in this case, it is still possible to perform the inner product estimation
presented in the main text. However, the computational complexity increases
according to the success probability of the post selection. Specifically,
it increases by a factor of $O(\eta\kappa^2)$ in the present case.

\subsection{Summary and concluding remarks}

The example of the reference state preparation presented in this section
demonstrates the existence of quantum circuits with polylogarithmic complexity
in terms of $N$, capable of generating $\ket{\chi_1}$ and $\ket{\chi_2}$
such that $1/\zeta = O(\operatorname{poly}R)$. The construction of these
circuits necessitates the pure-state tomography of right singular vector states.
Consequently, the computational complexity of constructing the presented circuits
increases linearly with $M$, while the execution of the presented circuits is
much more efficient with $O(\operatorname{poly} \log M)$ complexity.

It is important to note that while this example serves as a theoretical
construct to demonstrate that the qDMD algorithm can achieve an exponential
speedup in $N$, it is not optimized for practical use. For specific problems,
one may find more efficient ways to prepare reference states. For instance,
methods like the automatic quantum circuit encoding \cite{Shirakawa2021} may
construct quantum circuits that approximately generate singular vector states
or their superpositions without exhaustive tomography.
Furthermore, the value of $\zeta$ may be enhanced in a variational manner
in which a parameterized ansatz quantum circuit is tuned based on observed
inner product values. Such variational method may improve the scaling of
the computational complexity relative to $R$.
The development of a pragmatic approach to the reference state preparation
remains an open area for future research.

\section{Computing DMD Mode States}

\subsection{Recursive coherent state addition}

The quantum circuit proposed by Oszmaniec et al. \cite{Oszmaniec2016}
(Fig. 3 of the main text) creates a coherent superposition of
two quantum states $\ket{\psi_0}$ and $\ket{\psi_1}$:
\begin{equation}
\ket{\Psi} =
\alpha\frac{\braket{\chi|\psi_1}}{|\!\braket{\chi|\psi_1}\!|}\ket{\psi_0} +
\beta \frac{\braket{\chi|\psi_0}}{|\!\braket{\chi|\psi_0}\!|}\ket{\psi_1}.
\end{equation}
Here, $\alpha$ and $\beta$ are user-specified complex amplitudes, and
$\ket{\chi}$ is a reference quantum state for the coherent addition.
When $\ket{\psi_0}$ and $\ket{\psi_1}$ are orthogonal---the computation
of DMD mode states satisfies this condition---, the probability of
successfully creating the coherent superposition is given by
\begin{equation}
\frac{|\!\braket{\chi|\psi_0}\!|^2|\!\braket{\chi|\psi_1}\!|^2}
     {|\!\braket{\chi|\psi_0}\!|^2+|\!\braket{\chi|\psi_1}\!|^2}.
\end{equation}
The created superposition can also be represented as
\begin{equation}
\begin{split}
\ket{\Psi}
&= \alpha \mathrm{e}^{\mathrm{i} \theta_1}\ket{\psi_0} +
   \beta  \mathrm{e}^{\mathrm{i} \theta_0}\ket{\psi_1} \\
&= \mathrm{e}^{\mathrm{i} (\theta_0 + \theta_1)} \left[
     \alpha \mathrm{e}^{-\mathrm{i} \theta_0}\ket{\psi_0} +
     \beta  \mathrm{e}^{-\mathrm{i} \theta_1}\ket{\psi_1}
\right],
\end{split}
\end{equation}
where
\begin{equation}
\theta_i = \arg \braket{\chi|\psi_i}
\end{equation}
for $i=0, 1$. If the global phase of $\ket{\psi_i}$ is shifted by $\theta$,
the phase $\theta_i$ is also shifted by $\theta$. Therefore, the phase factor
$\exp(-\mathrm{i}\theta_i)$ of the amplitude of $\ket{\psi_i}$ ensures that
the superposition state $\ket{\Psi}$ is invariant up to its global phase
under a global phase shift of $\ket{\psi_i}$. When $\theta_0$ and $\theta_1$
are known, including the phase factors $\exp(\mathrm{i}\theta_0)$ and
$\exp(\mathrm{i}\theta_1)$ into user-specified amplitudes $\alpha$ and $\beta$,
one can create a superposition of two quantum states with arbitrary amplitudes.

\begin{figure}[!tb]
    \centering
    \includegraphics{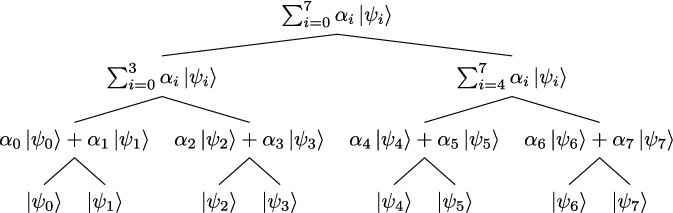}
    \caption{
      Recursive coherent state addition for creating a superposition of multiple
      quantum states. This figure illustrates the case for an eight-state
      superposition. The root node represents the target superposition of
      multiple quantum states. The leaf nodes correspond to elementary
      quantum states comprising the target superposition. Each branch node
      (a non-leaf node) represents a coherent superposition of its child nodes,
      up to normalization. The process advances from the bottom to the top,
      where the superposition at each branch node is created from the coherent
      addition of its child nodes.
    }
    \label{fig:coherent-state-addition-tree}
\end{figure}

One can also create a coherent superposition of multiple quantum states
by recursively creating two-state coherent superpositions.
This recursive coherent state addition process is illustrated in
Fig.~\ref{fig:coherent-state-addition-tree}. In this figure, the multi-state coherent
addition process proceeds from the bottom to the top; the superposition state
at each branch node is created by the two-state coherent addition of its child
nodes' states. One may adopt different reference states for each two-state
coherent addition process to enhance the success probability of each process
(see below).

The recursive coherent state addition protocol allows for the computation
of a DMD mode state. Suppose we aim to compute a DMD mode state
\begin{equation}
\ket{\tilde{\bm{w}}_r} \approx
\sum_{r^\prime=1}^R \tilde{w}_r^{\prime r^\prime}
\ket{\bm{u}^{\tiny [\bm{X}\ \bm{X}^\prime]}_{r^\prime}}.
\end{equation}
The recursive coherent state addition consists of $R-1$ steps of two-state
coherent addition in this case. Let us consider a two-state coherent addition
step indexed by $k$ that adds two quantum states
\begin{align}
\ket{\psi_{I_0^k}} &=
\frac{1}{\sqrt{\sum_{r^\prime \in I_0^k} |\tilde{w}_r^{\prime r^\prime}|^2}}
\sum_{r^\prime \in I_0^k} \tilde{w}_r^{\prime r^\prime}
\ket{\bm{u}^{\tiny [\bm{X}\ \bm{X}^\prime]}_{r^\prime}}, \\
\ket{\psi_{I_1^k}} &=
\frac{1}{\sqrt{\sum_{r^\prime \in I_1^k} |\tilde{w}_r^{\prime r^\prime}|^2}}
\sum_{r^\prime \in I_1^k} \tilde{w}_r^{\prime r^\prime}
\ket{\bm{u}^{\tiny [\bm{X}\ \bm{X}^\prime]}_{r^\prime}},
\end{align}
to create the superposition state
\begin{equation}
\ket{\psi_{I_0^k \cup I_1^k}} =
\frac{1}
     {\sqrt{\sum_{r^\prime \in I_0^k \cup I_1^k}
            |\tilde{w}_r^{\prime r^\prime}|^2}}
\sum_{r^\prime \in I_0^k \cup I_1^k} \tilde{w}_r^{\prime r^\prime}
\ket{\bm{u}^{\tiny [\bm{X}\ \bm{X}^\prime]}_{r^\prime}}.
\end{equation}
Here, $I_0^k$ and $I_1^k$ are some mutually-exclusive sets of indices
$r^\prime$. The user-specified amplitude of $\ket{\psi_{I_j^k}}$ ($j=0, 1$)
to create $\ket{\psi_{I_0^k \cup I_1^k}}$ is given by
\begin{equation}
\sqrt{\frac{\sum_{r^\prime \in I_j^k}|\tilde{w}_r^{\prime r^\prime}|^2}
     {\sum_{r^\prime \in I_0^k \cup I_1^k}|\tilde{w}_r^{\prime r^\prime}|^2}}
\mathrm{e}^{\mathrm{i}\theta_{I_j^k}},
\end{equation}
where
\begin{equation}
\theta_{I_j^k} = \arg \braket{\chi_\mathrm{add}^{k}|\psi_{I_j^k}},
\end{equation}
and $\ket{\chi_\mathrm{add}^{k}}$ denotes the reference state for
the two-state coherent addition of step $k$. The inner product
$\braket{\chi_\mathrm{add}^{k}|\psi_{I_j^k}}$ can be calculated
through the equation
\begin{equation}
\braket{\chi_\mathrm{add}^{k}|\psi_{I_j^k}} =
\frac{1}{\sqrt{\sum_{r^\prime \in I_j^k} |\tilde{w}_r^{\prime r^\prime}|^2}}
\sum_{r^\prime \in I_j^k} \tilde{w}_r^{\prime r^\prime}
\braket{\chi_\mathrm{add}^{k}|
        \bm{u}^{\tiny [\bm{X}\ \bm{X}^\prime]}_{r^\prime}}
\label{eq:inner-product-estimation-for-two-state-addition}
\end{equation}
and the estimation of
$\{\braket{\chi_\mathrm{add}^{k}|
           \bm{u}^{\tiny [\bm{X}\ \bm{X}^\prime]}_{r^\prime}}\}$.
To estimate
$\braket{\chi_\mathrm{add}^{k}|
         \bm{u}^{\tiny [\bm{X}\ \bm{X}^\prime]}_{r^\prime}}$,
we perform the three-state SWAP test for $\ket{\chi_\mathrm{add}^{k}}$,
$\ket{\bm{u}^{\tiny [\bm{X}\ \bm{X}^\prime]}_{r^\prime}}$, and
$\ket{\chi_1}$, which provides an estimate of
$\braket{\chi_\mathrm{add}^{k}|
         \bm{u}^{\tiny [\bm{X}\ \bm{X}^\prime]}_{r^\prime}}\!
 \braket{\bm{u}^{\tiny [\bm{X}\ \bm{X}^\prime]}_{r^\prime}|\chi_1}\!
 \braket{\chi_1|\chi_\mathrm{add}^{k}}$.
The inner product
$\braket{\bm{u}^{\tiny [\bm{X}\ \bm{X}^\prime]}_{r^\prime}|\chi_1}$
is known through the estimation of $\tilde{\bm{K}}^\prime$.
The inner product $\braket{\chi_1|\chi_\mathrm{add}^{k}}$ can be estimated
as follows. Recall that the reference states $\ket{\chi_1}$ and
$\ket{\chi_\mathrm{add}^{k}}$ are assumed to be prepared unitary gates
$U_{\chi_1}$ and $U_{\chi_\mathrm{add}^{k}}$, respectively\footnote{
  Instead of the assumption of the unitarity of the reference state
  preparation, the assumption of the availability of block encodings of
  $U_{\chi_1}$ and $U_{\chi_\mathrm{add}^{k}}$ also makes the argument
  in the present subsection valid, although the computational complexity
  may increase.
}. Using these unitary gates conditionally, we compute the following state:
\begin{equation}
\frac{U_{\chi_1}\ket{0}_1\!\ket{0}_\mathrm{a} +
      U_{\chi_\mathrm{add}^{k}}\ket{0}_1\!\ket{1}_\mathrm{a}}
     {\sqrt{2}} =
\frac{\ket{\chi_1}_1\!\ket{0}_\mathrm{a} +
      \ket{\chi_\mathrm{add}^{k}}_1\!\ket{1}_\mathrm{a}}
     {\sqrt{2}},
\end{equation}
where $\ket{\ }_\mathrm{a}$ denotes a ket of an ancilla control qubit.
When the above state is input to the circuit depicted in Fig. 2
of the main text, the circuit provides an estimate of
$\braket{\chi_1|\chi_\mathrm{add}^{k}}$. Therefore, we can estimate
the inner products
$\{\braket{\chi_\mathrm{add}^{k}|
           \bm{u}^{\tiny [\bm{X}\ \bm{X}^\prime]}_{r^\prime}}\}$
and calculate $\braket{\chi_\mathrm{add}^{k}|\psi_{I_j^k}}$.
In this way, we determine the user-specified amplitudes in each step $k$.
By repeating two-state coherent addition processes recursively
in the manner shown in Fig.~\ref{fig:coherent-state-addition-tree},
we get the desired DMD mode vector state.

\subsection{Computational complexity}

Let us evaluate the computational complexity of computing a DMD mode state
with keeping the phase error of each amplitude $\tilde{w}_r^{\prime r^\prime}$
below $\epsilon$. The computational complexity consists of two contributing
parts: (1) the computational complexity of estimating phases $\theta_{I_0^k}$
and $\theta_{I_1^k}$ at all steps $k$ and (2) the computational complexity of
the recursive coherent state addition. In the following evaluation,
we omit the computational costs for reference state preparations,
which will be considered in the next subsection.

First, we consider the computational complexity of estimating phases
$\theta_{I_0^k}$ and $\theta_{I_1^k}$ at all steps $k$.
Because the tree depth for the recursive coherent state addition is
$\lceil\log_2 R\rceil$, the tolerant phase error at each step is
$\epsilon/\lceil\log_2 R\rceil$. Therefore, at each step $k$,
it is necessary to estimate
$\braket{\chi_\mathrm{add}^{k}|\psi_{I_j^k}}$ with precision\footnote{
  Let $\tilde{z}$ be an estimate of a complex value $z$ and $\epsilon_z$
  denote the estimation error, i.e., $\tilde{z} = z + \epsilon_z$.
  Here, we assume that $|\epsilon_z| < |z|$. On the complex plane,
  when the absolute value of $\epsilon_z$ is fixed, $\tilde{z}$ lies on
  the circle centered at $z$ with radius $|\epsilon_z|$. The phase error
  $\epsilon_\mathrm{phase} = |\arg \tilde{z} - \arg z|$ takes the maximum
  when the line passing through $\tilde{z}$ and the origin is tangent to
  the circle. The maximum value of $\epsilon_\mathrm{phase}$ is
  $\arcsin(|\epsilon_z/z|)$. Because $\arcsin(|\epsilon_z/z|)$
  is less than $(\pi/2)|\epsilon_z/z|$, the estimation of $z$ with precision
  $(2/\pi)|z|\epsilon$ ensures that the phase error is less than $\epsilon$.
}
\begin{equation}
\frac{2}{\pi}
|\!\braket{\chi_\mathrm{add}^{k}|\psi_{I_j^k}}\!|
\frac{\epsilon}{\lceil\log_2 R\rceil}.
\end{equation}
The estimation error of
$\braket{\chi_\mathrm{add}^{k}|\psi_{I_j^k}}$
is bounded from above by the sum of the estimation errors of the terms on
the right hand side of Eq.~\eqref{eq:inner-product-estimation-for-two-state-addition}.
Let $\epsilon_\mathrm{term}$ denote the maximum error of
$\{\braket{\chi_\mathrm{add}^{k}|
           \bm{u}^{\tiny [\bm{X}\ \bm{X}^\prime]}_{r^\prime}}\}$.
Then, the estimation error of
$\braket{\chi_\mathrm{add}^{k}|\psi_{I_j^k}}$
is bounded from above by
\begin{equation}
\frac{\sum_{r^\prime \in I_j^k} |\tilde{w}_r^{\prime r^\prime}|}
     {\sqrt{\sum_{r^\prime \in I_j^k} |\tilde{w}_r^{\prime r^\prime}|^2}}
     \epsilon_\mathrm{term}
\le \sqrt{|I_j^k|} \epsilon_\mathrm{term}
\le \sqrt{R} \epsilon_\mathrm{term}.
\end{equation}
The first inequality follows the Cauchy--Schwarz inequality.
Consequently,
\begin{equation}
\epsilon_\mathrm{term} =
\frac{2}{\pi}
|\!\braket{\chi_\mathrm{add}^{k}|\psi_{I_j^k}}\!|
\frac{\epsilon}{\sqrt{R}\lceil\log_2 R\rceil}
\end{equation}
is a sufficient condition for the desired total precision.
This precision $\epsilon_\mathrm{term}$ is achieved when
$\braket{\bm{u}^{\tiny [\bm{X}\ \bm{X}^\prime]}_{r^\prime}|\chi_1}$,
$\braket{\chi_1|\chi_\mathrm{add}^{k}}$, and
$\braket{\chi_\mathrm{add}^{k}|
         \bm{u}^{\tiny [\bm{X}\ \bm{X}^\prime]}_{r^\prime}}\!
 \braket{\bm{u}^{\tiny [\bm{X}\ \bm{X}^\prime]}_{r^\prime}|\chi_1}\!
 \braket{\chi_1|\chi_\mathrm{add}^{k}}$ are estimated
with precision of the order of
\begin{equation}
\frac{2}{\pi}
|\!\braket{\chi_\mathrm{add}^{k}|\psi_{I_j^k}}\!
\braket{\bm{u}^{\tiny [\bm{X}\ \bm{X}^\prime]}_{r^\prime}|\chi_1}\!
\braket{\chi_1|\chi_\mathrm{add}^{k}}\!|
\frac{\epsilon}{\sqrt{R}\lceil\log_2 R\rceil}.
\end{equation}
Let us define parameters $\zeta_3$ and $\zeta_4$ as
\begin{align}
\zeta_3 &\coloneqq \min_k |\!\braket{\chi_1|\chi_\mathrm{add}^{k}}\!|^2, \\
\zeta_4 &\coloneqq \min_k \min_{j\in\{0, 1\}}
|\!\braket{\chi_\mathrm{add}^{k}|\psi_{I_j^k}}\!|^2,
\label{eq:zeta-def2}
\end{align}
where $\min_k$ signifies the minimum over all step-indices $k$. Additionally,
$\zeta_1$ defined in Eq.~\eqref{eq:zeta1-def} is a lower bound for
$|\!\braket{\bm{u}^{\tiny [\bm{X}\ \bm{X}^\prime]}_{r^\prime}|\chi_1}\!|^2$.
Using these parameters, we can express the required number of copies of each
$\ket{\bm{u}^{\tiny [\bm{X}\ \bm{X}^\prime]}_{r^\prime}}$
for estimating the phase factors at each step as
$O(R(\log_2 R)^2/\zeta_1\zeta_3\zeta_4\epsilon^2)$. Because each
$\ket{\bm{u}^{\tiny [\bm{X}\ \bm{X}^\prime]}_{r^\prime}}$ is involved
in $\lceil\log_2 R\rceil$ steps of two-state coherent addition,
the required number of quantum SVDs for estimating all phase factors is
\begin{equation}
O\left(\frac{\kappa_{\tiny [\bm{X}\ \bm{X}^\prime]}^2}
            {\zeta_1\zeta_3\zeta_4 \epsilon^2}
       R^2 (\log_2 R)^3
       \log(\kappa_{\tiny [\bm{X}\ \bm{X}^\prime]}^2 \log R)\right),
\end{equation}
due to Eq.~\eqref{eq:Brayton}. The parameter $\kappa_{\tiny [\bm{X}\ \bm{X}^\prime]}$
is the condition number of $[\bm{X}\ \bm{X}^\prime]$ (see Eq.~\eqref{eq:kappa-def}).
The gate complexity is larger than the required number of quantum SVDs
by a factor of $T/\epsilon^2\operatorname{poly} \log(NM/\epsilon)$.

Next, we consider the computational complexity of the recursive coherent
addition. The success probability at step $k$ can be bounded from below as
\begin{equation}
\frac{|\!\braket{\chi_\mathrm{add}^{k}|\psi_{I_0^k}}\!|^2
      |\!\braket{\chi_\mathrm{add}^{k}|\psi_{I_1^k}}\!|^2}
     {|\!\braket{\chi_\mathrm{add}^{k}|\psi_{I_0^k}}\!|^2 +
      |\!\braket{\chi_\mathrm{add}^{k}|\psi_{I_1^k}}\!|^2}
\ge \frac{1}{2}
    \min\{|\!\braket{\chi_\mathrm{add}^{k}|\psi_{I_0^k}}\!|^2,
          |\!\braket{\chi_\mathrm{add}^{k}|\psi_{I_1^k}}\!|^2\}
\ge \frac{\zeta_4}{2}.
\end{equation}
Therefore, $O(2/\zeta_4)$ copies of $\ket{\psi_{I_j^k}}$ ($j=0, 1$)
are required for successfully creating a superposition state at step $k$.
Because the tree depth for the recursive coherent state addition
is $\lceil\log_2 R\rceil < \log_2 R + 1$, the required number of copies of
$\ket{\bm{u}^{\tiny [\bm{X}\ \bm{X}^\prime]}_{r^\prime}}$ is
$O((2/\zeta_4)^{\log_2 R + 1})$  in total.
Due to Eq.~\eqref{eq:Brayton}, the required number of quantum SVDs for
the recursive coherent state addition is
\begin{equation}
O\left(\frac{\kappa_{\tiny [\bm{X}\ \bm{X}^\prime]}^2}{\zeta_4}
  R^{2 + \log_2 \frac{1}{\zeta_4}}
  \log(\kappa_{\tiny [\bm{X}\ \bm{X}^\prime]}^2 \log R)\right).
\end{equation}
The gate complexity is larger than the required number of quantum SVDs
by a factor of $T/\epsilon^2\operatorname{poly} \log(NM/\epsilon)$.

\subsection{Reference state preparation}

One may use any reference state at each step of the recursive coherent
state addition, provided that $\zeta_3$ and $\zeta_4$ are both positive.
Regardless of the choice of reference states, superposition states
to be computed are identical within a tolerant error. However,
the choice of reference states affects the computational complexity;
the larger $\zeta_3$ and $\zeta_4$ are, the shorter the computational time is.

A set of reference states with large $\zeta_3$ and $\zeta_4$ may be found by
a variational quantum algorithm. Let $U_k(\bm{\theta}_k)$ be an ansatz
quantum circuit for generating a reference state of step $k$:
$U_k(\bm{\theta}_k)\ket{0} = \ket{\chi_\mathrm{add}^{k}(\bm{\theta}_k)}$.
Here, $\bm{\theta}_k$ denotes circuit parameters to be optimized.
In this setup, $\zeta_3$ and $\zeta_4$ are considered as functions of
$\{\bm{\theta}_k\}$. The optimal values of $\{\bm{\theta}_k\}$ can be
found by solving the following optimization problem:
\begin{equation}
\operatorname*{maximize}_{\{\bm{\theta}_k\}}
\ \min\{\zeta_3(\{\bm{\theta}_k\}), \zeta_4(\{\bm{\theta}_k\})\}.
\end{equation}
Using a solution of the optimization problem, denoted by
$\{\bm{\theta}_k^*\}$, the reference state of each step $k$
is prepared by $U_k(\bm{\theta}_k^*)$.

We can also compute a set of reference states such that
$1/\zeta_3 = O(1)$ and $1/\zeta_4 = O(1)$ with another protocol.
These conditions on $\zeta_3$ and $\zeta_4$ imply that the required number of
quantum SVDs for computing a DMD mode state is $O(\operatorname{poly} R)$.
In what follows, we present a specific procedure for reference state
preparation satisfying these conditions.

Initially, we perform pure-state tomography \cite{Verdeil2023} to estimate
all right singular vector states of $[\bm{X}\ \bm{X}^\prime]$. The pure-state
tomography requires an $O(M)$ runtime for each right singular vector state.
Subsequently, for each right singular vector state
$\ket{\bm{v}^{\tiny [\bm{X}\ \bm{X}^\prime] *}_{r^\prime}}$,
we construct a quantum circuit $V_{r^\prime}$ such that
$V_{r^\prime}\ket{0} = \exp(-\mathrm{i}\varphi_{r^\prime})
 \!\ket{\bm{v}^{\tiny [\bm{X}\ \bm{X}^\prime] *}_{r^\prime}}$.
Here $\varphi_{r^\prime}$ signifies an unknown global phase which cannot be
determined by the pure-state tomography. When the classical data of
the amplitudes of $\ket{\bm{v}^{\tiny [\bm{X}\ \bm{X}^\prime] *}_{r^\prime}}$
is given, such a quantum circuit can be implemented with a gate complexity of
$O(\operatorname{poly}\log M)$ \cite{Kerenidis2017,Mitarai2019}.

Next, we show that a superposition state of the form
\begin{equation}
\sum_{r^\prime=1}^R \alpha_{r^\prime}
\mathrm{e}^{\mathrm{i}\varphi_{r^\prime}}
\ket{\bm{u}^{\tiny [\bm{X}\ \bm{X}^\prime]}_{r^\prime}}
\end{equation}
can be computed for arbitrary user-specified amplitudes
$\bm{\alpha} = [\alpha_1,\dots,\alpha_R]^\top$ such that $\|\bm{\alpha}\| = 1$,
using the quantum SVD and controlled-$V_{r^\prime}^\dagger$ operations.
The computation is composed of the following steps:
(1) Prepare $\ket{\mathrm{SVD}(\hat{[\bm{X}\ \bm{X}^\prime]})}_{1:5}$.
(2) Uncompute the second to fourth registers by
controlled-$V_{r^\prime}^\dagger$ operations conditioned on the fifth register:
\begin{equation}
\begin{split}
&\sum_{r^\prime=1}^R \hat{\sigma}^{\tiny [\bm{X}\ \bm{X}^\prime]}_{r^\prime}
\ket{\bm{u}^{\tiny [\bm{X}\ \bm{X}^\prime]}_{r^\prime}}_1\!
V_{r^\prime}^\dagger
\ket{\bm{v}^{\tiny [\bm{X}\ \bm{X}^\prime] *}_{r^\prime}}_{234}\!
\ket{(\hat{\sigma}^{\tiny [\bm{X}\ \bm{X}^\prime]}_{r^\prime})^2}_5 \\ =
& \sum_{r^\prime=1}^R \hat{\sigma}^{\tiny [\bm{X}\ \bm{X}^\prime]}_{r^\prime}
\mathrm{e}^{\mathrm{i}\varphi_{r^\prime}}
\ket{\bm{u}^{\tiny [\bm{X}\ \bm{X}^\prime]}_{r^\prime}}_1\!
\ket{0}_{234}\!
\ket{(\hat{\sigma}^{\tiny [\bm{X}\ \bm{X}^\prime]}_{r^\prime})^2}_5.
\end{split}
\end{equation}
(3) Append an ancilla qubit as the sixth register.
(4) Apply controlled rotations of the ancilla qubit conditioned on
the fifth register to yield
\begin{equation}
\sum_{r^\prime=1}^R \hat{\sigma}^{\tiny [\bm{X}\ \bm{X}^\prime]}_{r^\prime}
\mathrm{e}^{\mathrm{i}\varphi_{r^\prime}}
\ket{\bm{u}^{\tiny [\bm{X}\ \bm{X}^\prime]}_{r^\prime}}_1\!\ket{0}_{234}\!
\ket{(\hat{\sigma}^{\tiny [\bm{X}\ \bm{X}^\prime]}_{r^\prime})^2}_5
\left[
     \frac{a\alpha_{r^\prime}}
          {\hat{\sigma}^{\tiny [\bm{X}\ \bm{X}^\prime]}_{r^\prime}}
     \ket{0}_6 +
     \sqrt{1 - 
          \left|\frac{a\alpha_{r^\prime}}
          {\hat{\sigma}^{\tiny [\bm{X}\ \bm{X}^\prime]}_{r^\prime}}\right|^2}
     \ket{1}_6
\right],
\end{equation}
where
\begin{equation}
a = \min_{r^\prime}
    \left|\frac{\hat{\sigma}^{\tiny [\bm{X}\ \bm{X}^\prime]}_{r^\prime}}
               {\alpha_{r^\prime}}\right|.
\end{equation}
(5) Uncompute the fifth register by the inverse operation of the quantum SVD.
(6) Measure the second to sixth registers. If all measured values are zero,
we obtain the target superposition state. The success probability is $|a|^2$,
bounded from below as
\begin{equation}
|a|^2 \ge
\min_{r^\prime} (\hat{\sigma}^{\tiny [\bm{X}\ \bm{X}^\prime]}_{r^\prime})^2 =
\Omega\left(\frac{1}{\kappa_{\tiny [\bm{X}\ \bm{X}^\prime]}^2 R}\right).
\end{equation}
Here, $\kappa_{\tiny [\bm{X}\ \bm{X}^\prime]}^2$ is the condition number of
$[\bm{X}\ \bm{X}^\prime]$ (see Eq.~\eqref{eq:kappa-def}), and we use
Eq.~\eqref{eq:kappa-sigma_min-relation} to derive the lower bound.
Therefore, the required number of quantum SVDs for this computation is
$O(\kappa_{\tiny [\bm{X}\ \bm{X}^\prime]}^2 R)$.

It is worth noting that the target superposition state
created by the above protocol contains unknown phase factors
$\exp(\mathrm{i}\varphi_{r^\prime})$; thus the above protocol
cannot be utilized for computing DMD mode states.
However, this protocol has less computational complexity compared
with the recursive coherent state addition protocol and is helpful
for preparing reference states as described below. Hereinafter,
we call the above protocol as \textit{out-of-phase superposition protocol}.

The reference state $\ket{\chi_\mathrm{add}^{k}}$ of step $k$ that we aim to
compute using the out-of-phase superposition protocol is given by
\begin{equation}
\ket{\chi_\mathrm{add}^{k}} =
\frac{1}{\sqrt{2}}
\left[\mathrm{e}^{\mathrm{i}\varphi_0^k}\ket{\psi_{I_0^k}} +
      \mathrm{e}^{\mathrm{i}\varphi_1^k}\ket{\psi_{I_1^k}}\right],
\label{eq:ref-state-addition}
\end{equation}
where $\varphi_j^k$'s signify arbitrary relative phases. This reference
state of step $k$ is computable for any $k$, which is proven by
the following mathematical induction with respect to the recursion depth.
(1) For any step $k$ that corresponds to one of the deepest branch nodes of
the recursion tree (see Fig.~\ref{fig:coherent-state-addition-tree}),
$\ket{\psi_{I_j^k}}$'s are left singular vector states of
$[\bm{X}\ \bm{X}^\prime]$, thus $\ket{\chi_\mathrm{add}^{k}}$ defined above
is computable by the out-of-phase superposition protocol.
(2) For step $k$ that does not correspond to one of the deepest branch nodes,
assume that the reference states for all descendant nodes of step $k$
are computable. Let $k_\mathrm{c}$ denote the index of a child node of
step $k$. Due to the induction hypothesis, we can compute
$\ket{\psi_{I_0^{k_\mathrm{c}} \cup I_1^{k_\mathrm{c}}}}$
by recursively performing the two-state coherent addition steps of
the descendant nodes with the reference states defined
in Eq.~\eqref{eq:ref-state-addition}. Furthermore, we can compute the following
state with control parameter $\vartheta^{k_\mathrm{c}} \in [0, 2\pi)$
using the out-of-phase superposition protocol:
\begin{equation}
\begin{split}
\ket{\tilde{\psi}_{I_0^{k_\mathrm{c}} \cup I_1^{k_\mathrm{c}}}
     (\vartheta^{k_\mathrm{c}})} =
&\frac{\mathrm{e}^{\mathrm{i}\varphi_0^{k_\mathrm{c}}}}
     {\sqrt{\sum_{r^\prime \in I_0^{k_\mathrm{c}} \cup I_1^{k_\mathrm{c}}}
      |\tilde{w}_r^{\prime r^\prime}|^2}}
\sum_{r^\prime \in I_0^{k_\mathrm{c}}} \tilde{w}_r^{\prime r^\prime}
\ket{\bm{u}^{\tiny [\bm{X}\ \bm{X}^\prime]}_{r^\prime}} \\
&+
\frac{\mathrm{e}^{\mathrm{i}(\varphi_1^{k_\mathrm{c}}
                  +\vartheta^{k_\mathrm{c}})}}
     {\sqrt{\sum_{r^\prime \in I_0^{k_\mathrm{c}} \cup I_1^{k_\mathrm{c}}}
      |\tilde{w}_r^{\prime r^\prime}|^2}}
\sum_{r^\prime \in I_1^{k_\mathrm{c}}} \tilde{w}_r^{\prime r^\prime}
\ket{\bm{u}^{\tiny [\bm{X}\ \bm{X}^\prime]}_{r^\prime}}.
\end{split}
\end{equation}
The user-specified amplitudes for computing this state by the out-of-phase
superposition protocol is given by
\begin{equation}
\bm{\alpha}^{k_\mathrm{c}}(\vartheta^{k_\mathrm{c}}) =
\sqrt{
\frac{2\sum_{r^\prime \in I_0^{k_\mathrm{c}}}
      |\tilde{w}_r^{\prime r^\prime}|^2}
     {\sum_{r^\prime \in I_0^{k_\mathrm{c}} \cup I_1^{k_\mathrm{c}}}
      |\tilde{w}_r^{\prime r^\prime}|^2}
}\bm{\alpha}_0^{k_\mathrm{c}} +
\mathrm{e}^{\mathrm{i}\vartheta^{k_\mathrm{c}}}\sqrt{
\frac{2\sum_{r^\prime \in I_1^{k_\mathrm{c}}}
      |\tilde{w}_r^{\prime r^\prime}|^2}
     {\sum_{r^\prime \in I_0^{k_\mathrm{c}} \cup I_1^{k_\mathrm{c}}}
      |\tilde{w}_r^{\prime r^\prime}|^2}
}\bm{\alpha}_1^{k_\mathrm{c}}.
\end{equation}
Here, $\bm{\alpha}_j^{k_\mathrm{c}}$ is a vector whose
$r^\prime$-th element is the user-specified amplitude of
$\ket{\bm{u}^{\tiny [\bm{X}\ \bm{X}^\prime]}_{r^\prime}}$
used to compute $\ket{\chi_\mathrm{add}^{k_\mathrm{c}}}$
if $r^\prime \in I_j^{k_\mathrm{c}}$ or zero otherwise.
Note that $\bm{\alpha}_j^{k_\mathrm{c}}$'s are known due to
the induction hypothesis. This state
$\ket{\tilde{\psi}_{I_0^{k_\mathrm{c}} \cup I_1^{k_\mathrm{c}}}
 (\vartheta^{k_\mathrm{c}})}$ is similar to the state
$\ket{\psi_{I_0^{k_\mathrm{c}} \cup I_1^{k_\mathrm{c}}}}$.
The fidelity of these two states, i.e.,
$|\!\braket{\tilde{\psi}_{I_0^{k_\mathrm{c}} \cup I_1^{k_\mathrm{c}}}
            (\vartheta^{k_\mathrm{c}})|
            \psi_{I_0^{k_\mathrm{c}} \cup I_1^{k_\mathrm{c}}}}\!|^2$,
takes a maximum value of one if and only if
$\vartheta^{k_\mathrm{c}} = \varphi_0^{k_\mathrm{c}}-\varphi_1^{k_\mathrm{c}}$.
The maximizer $\vartheta^{k_\mathrm{c}*}$ of the fidelity can be found
through a grid search based on the fidelity measurement. The quantum state
$\ket{\tilde{\psi}_{I_0^{k_\mathrm{c}} \cup I_1^{k_\mathrm{c}}}
      (\vartheta^{k_\mathrm{c} *})}$
is equivalent to
$\ket{\psi_{I_0^{k_\mathrm{c}} \cup I_1^{k_\mathrm{c}}}}
 \ (=\ket{\psi_{I_j^k}})$
up to a global phase. In this way, we can compute the user-specified amplitudes
$\bm{\alpha}^{k_\mathrm{c}}(\vartheta^{k_\mathrm{c} *})$ to compute
$\ket{\psi_{I_j^k}}$ by the out-of-phase superposition protocol. By employing
$\sum_{k_\mathrm{c}}
 \bm{\alpha}^{k_\mathrm{c}}(\vartheta^{k_\mathrm{c} *})/\sqrt{2}$
as a user-specified amplitude vector, we are able to compute the reference
state of the form given in Eq.~\eqref{eq:ref-state-addition} using the out-of-phase
superposition protocol. This concludes the proof.

The above proof shows that we can recursively determine an appropriate
amplitude vector to compute $\ket{\psi_{I_j^k}}$ using the out-of-phase
superposition protocol. Therefore, $\ket{\psi_{I_0^k \cup I_1^k}}$ is
computable by a \textit{single-step} coherent addition of the two states
$\ket{\psi_{I_0^k}}$ and $\ket{\psi_{I_1^k}}$, which are prepared with
the out-of-phase superposition protocol. This quantum process has
an advantage in the success probability compared with the \textit{multistep}
recursive coherent addition, because the success probability of
the recursive coherent addition decreases exponentially with respect to
the recursion depth. In addition, the presented quantum process
for computing $\ket{\chi_\mathrm{add}^k}$ is not unitary
but a block encoding of the unitary gate $U_{\chi_\mathrm{add}^k}$.
Even in this case, it is still possible to perform the coherent addition,
although the success probability decreases by a factor of
$\Omega(1/\kappa_{\tiny [\bm{X}\ \bm{X}^\prime]}^2 R)$.
Given these facts, the necessary number of quantum SVDs to compute
$\ket{\psi_{I_0^k \cup I_1^k}}$ by the single-step coherent addition is
estimated as $O(\kappa_{\tiny [\bm{X}\ \bm{X}^\prime]}^4 R^2/\zeta_4)$.

The reference states defined in Eq.~\eqref{eq:ref-state-addition} satisfy
\begin{equation}
|\!\braket{\chi_\mathrm{add}^{k}|\psi_{I_j^k}}\!|^2 = \frac{1}{2}
\end{equation}
for all $k$ and $j$; thus $1/\zeta_4 = 2$.
In contrast, $|\!\braket{\chi_1|\chi_\mathrm{add}^{k}}\!|$ could be
close to zero for some $k$, leading to a situation where $1/\zeta_3$ becomes 
excessively large. This problem can be resolved by the following modification:
Let $b$ be a positive constant less than $1/\sqrt{2}$.
When the reference state satisfies
$|\!\braket{\chi_1|\chi_\mathrm{add}^{k}}\!| \le b/2$,
we modify the reference state to $\ket{\chi_\mathrm{add^\prime}^{k}}$:
\begin{equation}
\ket{\chi_\mathrm{add^\prime}^{k}} =
\mathcal{C}_\mathrm{add^\prime}^{k}
\left[\ket{\chi_\mathrm{add}^{k}} + b\ket{\chi_1}\right],
\end{equation}
where $\mathcal{C}_\mathrm{add^\prime}^{k}$ denotes the normalizing constant.
The normalizing constant is bounded from below as
\begin{equation}
\begin{split}
\mathcal{C}_\mathrm{add^\prime}^{k}
&= \frac{1}{\|\!\ket{\chi_\mathrm{add}^{k}} + b\ket{\chi_1}\!\|} \\
&\ge \frac{1}{\|\!\ket{\chi_\mathrm{add}^{k}}\!\|+b\|\ket{\chi_1}\!\|} =
\frac{1}{1+b}.
\end{split}
\end{equation}
Therefore, the modified reference state satisfies
\begin{equation}
\begin{split}
|\!\braket{\chi_1|\chi_\mathrm{add^\prime}^{k}}\!|
&\ge \mathcal{C}_\mathrm{add^\prime}^{k}
\left[b|\!\braket{\chi_1|\chi_1}\!| -
      |\!\braket{\chi_1|\chi_\mathrm{add}^{k}}\!|\right] \\
&\ge \frac{1}{1+b}\left(b - \frac{b}{2}\right) = \frac{b}{2(1+b)},
\end{split}
\end{equation}
and
\begin{equation}
\begin{split}
|\!\braket{\chi_\mathrm{add^\prime}^{k}|\psi_{I_j^k}}\!|
&\ge \mathcal{C}_\mathrm{add^\prime}^{k}
\left[|\!\braket{\chi_\mathrm{add}^{k}|\psi_{I_j^k}}\!| -
      b|\!\braket{\chi_1|\psi_{I_j^k}}\!|\right] \\
&\ge \frac{1}{1+b}\left(\frac{1}{\sqrt{2}}-b\right) > 0.
\end{split}
\end{equation}
Consequently, $1/\zeta_3 = O(1)$ and $1/\zeta_4 = O(1)$.

In summary, the overall number of quantum SVDs necessary for computing
a DMD mode state through the presented protocol scales as $\tilde{O}(R^2)$
with respect to $R$, although the protocol requires the pure state tomography
with $O(MR)$ complexity. Here, $\tilde{O}$ indicates that polylogarithmic
factors are omitted. The gate complexity is larger than the required number of
quantum SVDs by a factor of $T/\epsilon^2\operatorname{poly}\log(NM/\epsilon)$.

\section{Comparison of the \lowercase{q}DMD and Related Algorithms}

This section compares the qDMD algorithm proposed in the present study
with the related algorithms presented in previous studies
\cite{Steffens2017,Xue2023,Wang2010,Daskin2014,Teplukhin2020,Shao2022}.

\subsection{Time Series Analysis Algorithms}

First, we compare the qDMD algorithm with two quantum algorithms for
time series analysis, namely, (1) quantum matrix pencil method (QMPM) proposed
by Steffens et al. \cite{Steffens2017} and (2) a quantum dynamic mode decomposition
(QDMD\footnote{
   Xue et al. named their algorithm``QDMD," while we call our algorithm
  ``qDMD" because the DMD community sometimes prefixes the name of variants
  with lowercase letters. To avoid confusion, we also refer to the QDMD
  algorithm as``QRAM-based QDMD" in the text because the algorithm requires
  quantum random access memory (QRAM).
}) algorithm proposed by Xue et al. \cite{Xue2023}.
Table~\ref{tbl:time-series-analysis-algorithms} provides a summary of this comparison.

\begin{table*}[b]
  \centering
  \caption{Comparison of three quantum algorithms for time series analysis.}
  {\renewcommand\arraystretch{2}
  \begin{tabular}{p{2.5cm}|p{4.5cm}|l|l}
    \hline
    Algorithm & Time Series Data Input Method &
    Matrix Estimation Complexity & Mode State Computation Complexity \\
    \hline
    qDMD \quad (present study) & Quantum states computed by a quantum differential equation solver &
    $\tilde{O}\left(\frac{\kappa^2}{\zeta^2}\frac{R^2 T}{\epsilon^4}
                    \operatorname{poly}\log NM\right)$ &
    $\tilde{O}\left(\frac{\kappa^2}{\zeta_4}
                    \frac{R^{2+\log_2(1/\zeta_4)}T}{ \epsilon^2}
                    \operatorname{poly}\log NM\right)$ \\
    QMPM \cite{Steffens2017} & Quantum oracles that have random access to time series data &
    $\tilde{O}\left(\frac{\xi_\mathrm{QMPM}R^2}{\epsilon^4}\operatorname{poly}\log T\right)$ &
    N/A \\
    QDMD \cite{Xue2023} & Quantum oracles that have random access to time series data &
    $\tilde{O}\left(\frac{\kappa^5 R^6 M^{3.5}}{\epsilon^3}\operatorname{poly}\log NM\right)$ &
    $\tilde{O}\left(\kappa M\operatorname{poly}\log NM\right)$ \\
    \hline
    \end{tabular}}
  \label{tbl:time-series-analysis-algorithms}
\end{table*}

All the time series analysis algorithms aim to decompose time series data
into modes with exponential decay/growth and/or sinusoidal oscillation.
The qDMD and QDMD algorithms analyze $N$-dimensional time series data
with length $M$, while the QMPM algorithm analyzes one-dimensional data
with length $T$. All the algorithms are based on similar quantum-classical
hybrid strategies which consist of (i) estimating a matrix such as
$\tilde{\bm{K}}^\prime$ using a quantum computer, (ii) solving the eigenvalue
problem of the estimated matrix on a classical computer, and (iii) computing
mode states on a quantum computer. Since the QMPM algorithm analyzes
one-dimensional data, it does not include the mode state computing step.

The main difference between these algorithms lies in their input data.
The input to the qDMD algorithm is quantum states computed by a quantum
differential equation solver. On the other hand, the QMPM and QDMD algorithms
require quantum oracles\footnote{
  Steffens et al. also proposed another version of the QMPM algorithm
  where the input is quantum states encoding time series data in a particular
  form. However, how to prepare such particular quantum states remains an open
  question, and the main focus of the study is on the oracular setting
  \cite{Steffens2017}.
} that have random access to, for example, a time series data element $x_j^i$.
The implementation of such quantum random access memories (QRAMs) presents
technical challenges, as also highlighted in \cite{Kiani2022}. A QRAM may need to
access a classical data structure that stores the time series data \cite{Kerenidis2017}.
Given that the time series data comprises $NM$ entries in this study, constructing
such a QRAM may require computation time and memory of the order of $O(NM)$.
Therefore, these QRAM-based algorithms are likely not suitable for analyzing
high-dimensional differential equations when $N$ is too large for
classical computers to handle. Additionally, the QMPM algorithm is
inherently designed for one-dimensional time series data and does not
address the high-dimensional cases considered here.

The computational complexity of matrix estimation (i) and that of
mode state computation (iii) for each algorithm are listed in
Table~\ref{tbl:time-series-analysis-algorithms}. In the table, we omit
polylogarithmic factors in the computational complexities by use of
the symbol $\tilde{O}$, except for the data size parameters $N$, $M$,
and $T$. The parameter $\kappa$ denotes the maximum condition number
among matrices $\bm{X}$, $\bm{X}^\prime$, and $[\bm{X}\ \bm{X}^\prime]$
(see Eq.~\eqref{eq:kappa-def}). The parameters $\zeta$ and $\zeta_4$ are defined
in Eqs.~\eqref{eq:zeta-def} and \eqref{eq:zeta-def2}, respectively. The parameter
$\xi_\mathrm{QMPM}$ is related to sampling efficiency in the QMPM algorithm
\cite{Steffens2017}.

All algorithms achieve exponential speedups in the dimensionality $N$
and/or the time series data length $T$ or $M$. However, the computational
complexities of the QRAM-based algorithms listed in the table \textit{do not}
include the cost of time series data computation, while the computational
complexity of the qDMD algorithm \textit{does} include it. As noted above,
when the time series computation cost is taken into account,
the computational complexities of the QRAM-based algorithms may be
exponentially larger than those listed in the table with respect to
$T$ (for QMPM) or $N$ (for QDMD). In regard to other parameters,
the QRAM-based QDMD algorithm is better than the others in terms
of $\epsilon$, but our qDMD algorithm outperforms the QRAM-based
QDMD algorithm in terms of $M$.

\subsection{Quantum Eigensolvers for Complex Eigenvalue Problems}

Second, we compare the qDMD algorithm with four quantum eigensolvers for
complex eigenvalue problems, namely, (1) measurement-based phase estimation
algorithm (MPEA) proposed by Wang et al. \cite{Wang2010}, (2) iterative phase
estimation algorithm with universal circuit for non-unitary matrices
(IPEA+UCNUM) proposed by Daskin et al. \cite{Daskin2014}, (3) quantum annealer
eigensolver (QAE) proposed by Teplukhin et al. \cite{Teplukhin2020}, and
(4) phase estimation algorithm with a quantum linear differential equation
solver (PEA+QLDES) proposed by Shao \cite{Shao2022}.
Table~\ref{tbl:quantum-eigensolvers} provides a summary of this comparison.

\begin{table*}[b]
  \centering
  \caption{Comparison of quantum eigensolvers for complex eigenvalue problems.
           In the table, $\bm{A}$ denotes the $N \times N$ target matrix of
           an eigenvalue problem, and $\hat{A}$ signifies the operator form of $\bm{A}$.}
  {\renewcommand\arraystretch{2}
  \begin{tabular}{p{3cm}|p{8cm}|p{4cm}|p{2cm}}
    \hline
    Algorithm & Requirements &
    Outputs & Exponential Speedup in $N$ \\
    \hline
    qDMD \qquad \quad (present study) &
    (1) $\dot{\bm{x}}=\bm{A}\bm{x}$ can be simulated efficiently on a quantum computer and
    (2) the number of dominant eigenstates in the initial state is $O(\operatorname{poly}\log N)$. &
    Eigenvalues and eigenstates & YES \\
    MPEA \cite{Wang2010} &
    There exist a pure state $\ket{\phi}$ of an ancillary system, a Hamiltonian operator
    $\hat{H}$ acting on the target and ancillary systems, and a positive constant $\Delta t$
    such that $\hat{A}=\bra{\phi}\exp(-\mathrm{i}\hat{H}\Delta t)\ket{\phi}$. &
    The eigenvalue with the largest absolute value and its eigenstate & YES \\
    IPEA+UCNUM \cite{Daskin2014} &  An eigenstate can be prepared. &
    The eigenvalue of the prepared state & NO \\
    QAE \cite{Teplukhin2020} & $\bm{A}$ is Hermitian or symmetric. &
    Eigenvalues and eigenvectors (classical data) & NO \\
    PEA+QLDES \cite{Shao2022} &
    (1) $\dot{\bm{y}}=\mathrm{i}\pi(\bm{A}\otimes\bm{I}+\bm{I}\otimes\bm{A}^*)\bm{y}$ can be
    simulated efficiently and (2) an initial state $\bm{y}_0=\sum_i c_i(\bm{v}_i\otimes\bm{v}_i^*)$
    can be prepared efficiently on a quantum computer ($\bm{v}_i$: the $i$-th eigenvector,
    $c_i$: a complex constant, $^*$: complex conjugate). &
    Eigenvalues and eigenstates & YES \\
    \hline
    \end{tabular}}
  \label{tbl:quantum-eigensolvers}
\end{table*}

The first algorithm, MPEA, is a quantum analog of the power method
for solving eigenvalue problems. The algorithm iteratively applies
a non-unitary operator $\hat{A}$ to a state vector of the target system.
The operator $\hat{A}$ is realized through Hamiltonian simulation and
a projective measurement as
$\hat{A}=\bra{\phi}\exp(-\mathrm{i}\hat{H}\Delta t)\ket{\phi}$.
Here, $\ket{\phi}$ is a pure state of an ancillary system, $\hat{H}$ denotes
the Hamiltonian governing the dynamics of the target and ancillary systems,
and $\Delta t$ signifies the time step parameter. After applying $\hat{A}$
to an initial state $\ket{\psi_0}$ a sufficient number of times, one can obtain
the eigenstate of $\hat{A}$ with the maximum eigenvalue (in absolute value)
among all eigenstates included in $\ket{\psi_0}$. The MPEA is expected to
achieve an exponential speedup in the dimensionality of the target system $N$.
However, the success probability of the power iteration may decrease
exponentially as the number of iterations increases, leading to a decrease
in the algorithm's efficiency.

The second method applies the iterative phase estimation to the non-unitary
time evolution realized through a universal circuit design for arbitrary
non-unitary operations. This algorithm estimates the complex eigenvalue
associated with a prepared eigenstate. Because the complexity of
the non-unitary operation is $O(N^2)$, this algorithm does not
guarantee an exponential speedup in $N$.

The third algorithm, QAE, is a variational algorithm on quantum annealing
devices. This method can estimate complex eigenvalues and their eigenvectors
of complex Hermitian or symmetric matrices. Note that this algorithm outputs
eigenvectors as classical data instead of quantum states. This method also
does not guarantee an exponential speedup in $N$.

The fourth algorithm applies the phase estimation method to time series data
simulated by a QLDES. The algorithm simulates the linear differential
equation, defined as
$\dot{\bm{y}}=\mathrm{i}\pi(\bm{A}\otimes\bm{I}+\bm{I}\otimes\bm{A}^*)\bm{y}$.
Here, $\bm{A}^*$ designates the complex conjugate of $\bm{A}$.
The initial state of the simulation must be
$\bm{y}_0 = \sum_i c_i (\bm{v}_i \otimes \bm{v}_i^*)$,
where $\bm{v}_i$ and $\bm{v}_i^*$ denote the $i$-th eigenvector and
its complex conjugate, respectively, and $c_i$ is an arbitrary complex
constant. This algorithm can compute complex eigenvalues and
their eigenstates in time $O(\operatorname{poly}\log N)$.
However, it remains an open problem how to prepare the initial state of
the particular form without knowledge of the eigenstates.
Therefore, the author concluded that this algorithm is impractical
for complex eigenvalue problems \cite{Shao2022}.

The proposed qDMD algorithm requires that (1) $\dot{\bm{x}}=\bm{A}\bm{x}$ can
be simulated efficiently on a quantum computer and (2) the number of dominant
eigenstates in the initial state, i.e., $R$, is $O(\operatorname{poly}\log N)$.
These requirements are analogous to those in quantum phase estimation algorithm
for the eigenvalue problem of a Hermitian matrix $\bm{H}$ \cite{Abrams1999}:
(1) $\dot{\bm{x}}=-\mathrm{i}\bm{H}\bm{x}$ can be simulated efficiently on
a quantum computer and (2) the initial state includes
$O(\operatorname{poly}\log N)$ dominant eigenstates of which eigenvalues
are to be estimated. Under these requirements, the qDMD algorithm provides
an exponential speedup with respect to $N$ for finding complex eigenvalues
and eigenvectors.

\section{Dynamic Mode Decomposition for Defective Systems}

The qDMD algorithm presented in the main text assumes that the matrix
$\bm{A}$ is diagonalizable. This assumption implies that $\bm{K}$ is
also diagonalizable, thus its projected approximation $\tilde{\bm{K}}^\prime$
is assumed to have the eigenvalue decomposition. In this section, we consider
a situation where $\bm{A}$ is not diagonalizable. Such a non-diagonalizable
matrix is said to be \textit{defective} \cite{Golub1976}.

Any square matrix $\bm{A} \in \mathbb{C}^{N \times N}$ can be expressed
in the Jordan normal form \cite{Golub1976} as
\begin{equation}
\bm{A} =
\bm{P}
\begin{pmatrix}
\bm{J}_1 &        & \\
         & \ddots & \\
         &        & \bm{J}_p
\end{pmatrix}
\bm{P}^{-1},
\end{equation}
where $\bm{P}$ is an $N \times N$ invertible matrix, $\bm{J}_k$ ($k=1,\dots,p$)
denotes a Jordan block. The $k$-th Jordan block is defined by
\begin{equation}
\bm{J}_k =
\begin{pmatrix}
\lambda^{\tiny \bm{A}}_k & 1                        &        & \\
                         & \lambda^{\tiny \bm{A}}_k & \ddots & \\
                         &                          & \ddots & 1 \\
                         &                          &        & \lambda^{\tiny \bm{A}}_k
\end{pmatrix}
\in \mathbb{C}^{m_k \times m_k},
\end{equation}
where $\lambda^{\tiny \bm{A}}_k$ is an eigenvalue of $\bm{A}$,
and $m_k$ is the order of the Jordan block. The block diagonal matrix
$\operatorname{diag}(\bm{J}_1, \dots, \bm{J}_p)$ is called as
the Jordan normal form of $\bm{A}$. If $m_k = 1$ for all $k \in \{1,\dots,p\}$,
the Jordan normal form is a diagonal matrix, thus $\bm{A}$ is diagonalizable;
otherwise, $\bm{A}$ is defective. Using the Jordan normal form,
we can calculate the time-evolution operator $\bm{K}$ as
\begin{equation}
\begin{split}
\bm{K} &= \exp(\Delta t \bm{A}) \\
&= \bm{P}
\begin{pmatrix}
\exp(\Delta t \bm{J}_1) &        & \\
                        & \ddots & \\
                        &        & \exp(\Delta t \bm{J}_p)
\end{pmatrix}
\bm{P}^{-1}.
\end{split}
\end{equation}
The exponential of a Jordan block is given by
\begin{equation}
\exp(\Delta t \bm{J}_k) =
\mathrm{e}^{\Delta t \lambda^{\tiny \bm{A}}_k}
\begin{pmatrix}
1 & \frac{\Delta t}{1!} & \cdots & \frac{(\Delta t)^{m_k-1}}{(m_k-1)!} \\
  & 1                   & \ddots & \vdots \\
  &                     & \ddots & \frac{\Delta t}{1!}\\
  &                     &        & 1
\end{pmatrix}
\in \mathbb{C}^{m_k \times m_k}.
\end{equation}
For $\Delta t \neq 0$, this exponential $\exp(\Delta t \bm{J}_k)$ has
only one eigenvalue $\exp(\Delta t \lambda^{\tiny \bm{A}}_k)$ with
geometric multiplicity one. Therefore, the Jordan normal form
of $\bm{K}$ is the same as that of $\bm{A}$ except that
each eigenvalue $\lambda^{\tiny \bm{A}}_k$ is replaced by
$\exp(\Delta t \lambda^{\tiny \bm{A}}_k)$.

The qDMD algorithm, as well as the classical exact DMD algorithm \cite{Tu2014},
can estimate the time-evolution operator $\bm{K}$ through
$\tilde{\bm{K}}^\prime$ even for a defective system. This is because
$\tilde{\bm{K}}^\prime$ is defined through the singular value decompositions
of data matrices, where no assumption is made on the diagonalizablity.
However, for a defective system, $\tilde{\bm{K}}^\prime$ is (nearly)
defective\footnote{
  The Jordan normal form of $\bm{K}$ may be structurally instable;
  small perturbations due to finite computational precision and
  estimation errors may make $\tilde{\bm{K}}$ not strictly defective.
  In general, it is impossible to determine whether a matrix is defective
  or not in the presence of computational errors \cite{Golub1976}.
}, thus the eigenvalue decomposition of $\tilde{\bm{K}}^\prime$ is
infeasible. The Schur-based DMD algorithm proposed by Thitsa et al.
\cite{Thitsa2020} is a numerically-stable DMD algorithm for such
a (nearly) defective system. The Schur-based DMD algorithm computes
the Schur decomposition of $\tilde{\bm{K}}^\prime$, instead of
its eigenvalue decomposition. The Schur decomposition of
$\tilde{\bm{K}}^\prime$ is given by
\begin{equation}
\tilde{\bm{K}}^\prime =
\tilde{\bm{W}}_\mathrm{Schur}^\prime
\ \tilde{\bm{S}}^\prime
\ \tilde{\bm{W}}_\mathrm{Schur}^{\prime \dagger},
\end{equation}
where $\tilde{\bm{S}}^\prime$ is an upper triangular matrix,
and $\tilde{\bm{W}}_\mathrm{Schur}^\prime$ is a unitary matrix.
This decomposition provides an approximation of the Schur decomposition of
$\tilde{\bm{K}}$,
\begin{equation}
\tilde{\bm{K}} =
\tilde{\bm{W}}_\mathrm{Schur}
\ \tilde{\bm{S}}
\ \tilde{\bm{W}}_\mathrm{Schur}^{\dagger},
\end{equation}
as
\begin{align}
&\tilde{\bm{S}} \approx \tilde{\bm{S}}^\prime, \\
&\tilde{\bm{W}}_\mathrm{Schur} \approx
 \bm{Q} \tilde{\bm{W}}_\mathrm{Schur}^\prime.
\end{align}
Here, $\bm{Q}$ is an $N \times R$ matrix whose columns are
the $R$ dominant left singular vectors of $[\bm{X}\ \bm{X}^\prime]$.
The column vectors of the transformation matrix
$\tilde{\bm{W}}_\mathrm{Schur}$ can be computed
in the same manner as step 5 of the qDMD algorithm presented in
the main text. Therefore, the quantum procedures (steps 1--3 and 5)
of the qDMD algorithm can be directly adapted to the quantum version
of the Schur-based DMD algorithm for nearly defective systems.

\bibliography{supplement}